\documentclass[12pt]{iopart}
\usepackage[T1]{fontenc}
\usepackage[latin9]{inputenc}
\usepackage{amstext}
\usepackage{amssymb}
\usepackage{graphicx}
\usepackage{esint}

\makeatletter
\usepackage{iopams}
\usepackage{setstack}

\usepackage{color}
\usepackage{amstext}
\usepackage{iopams}

\makeatother

\begin{document}
\title[Efficient Quantum Jump Method for EET Dynamics]{An Efficient Quantum Jump Method for Coherent Energy Transfer Dynamics in Photosynthetic Systems under the Influence of Laser Fields}

\author{Qing Ai, Yuan-Jia Fan, Bih-Yaw Jin and Yuan-Chung Cheng\footnote{Author to whom correspondence should be addressed.}}

\address{Department of Chemistry and Center for Quantum Science and Engineering,
National Taiwan University, Taipei City 106, Taiwan}

\ead{yuanchung@ntu.edu.tw}
\begin{abstract}
We present a non-Markovian quantum jump approach for simulating coherent
energy transfer dynamics in molecular systems in the presence of laser
fields. By combining a coherent modified Redfield theory (CMRT) and
a non-Markovian quantum jump (NMQJ) method, this new approach inherits
the broad-range validity from the CMRT and highly efficient propagation
from the NMQJ. To implement NMQJ propagation of CMRT, we show that
the CMRT master equation can be casted into a generalized Lindblad
form. Moreover, we extend the NMQJ approach to treat time-dependent
Hamiltonian, enabling the description of excitonic systems under coherent
laser fields. As a benchmark of the validity of this new method, we
show that the CMRT-NMQJ method accurately describes the energy transfer
dynamics in a prototypical photosynthetic complex. Finally, we apply
this new approach to simulate the quantum dynamics of a dimer system
coherently excited to coupled single-excitation states under the influence
of laser fields, which allows us to investigate the interplay between
the photoexcitation process and ultrafast energy transfer dynamics
in the system. We demonstrate that laser-field parameters significantly
affect coherence dynamics of photoexcitations in excitonic systems,
which indicates that the photoexcitation process must be explicitly
considered in order to properly describe photon-induced dynamics in
photosynthetic systems. This work should provide a valuable tool for
efficient simulations of coherent control of energy flow in photosynthetic
systems and artificial optoelectronic materials.
\end{abstract}


\submitto{\NJP}

\maketitle

\section{Introduction}

During the past decade, much progress has been achieved in both experimental
and theoretical explorations of photosynthetic excitation energy transfer
(EET), e.g., EET pathways determined by pump-probe as well as two-dimensional
electronic spectroscopy (2DES) \cite{Brixner:2005wu,SchlauCohen:2009p77451,Novoderezhkin:2010p82122},
coherent EET dynamics revealed by 2DES \cite{Engel:2007hb,Collini:2010p79587,Panitchayangkoon:2010p82449},
and theoretical studies to elucidate mechanisms of EET \cite{Cheng:2009p75665,Cao:2009cc,OlayaCastro:2011hw,Jang:2012fi}.
It is intriguing to consider the possibility of using laser pulse
to coherently control energy flow in photosynthetic complexes. Notably,
coherent light sources were adopted to control energy transfer pathways
in LH2 \cite{Herek:2002dy}. The phases of the laser field can efficiently
adjust the ratio of energy transfer between intra- and inter-molecule
channels in the complex's donor-acceptor system. In addition, theoretical
control schemes were put forward to prepare specified initial states
and probe the subsequent dynamics \cite{Caruso:2012hv}, and the optimal
control theory has been adopted to optimize the effects of electromagnetic
field's polarization and structural and energetic disorder on localizing
excitation energy at a certain chromophore within a Fenna-Matthews-Olson
complex of green bacteria \cite{Bruggemann:2006ih}.

However, although the breakthroughs in experimental techniques have
provided much insight into the coherent dynamics in small photosynthetic
systems \cite{Ishizaki:2012kf,Dawlaty:2012fs}, efficient theoretical
methods that can describe quantum dynamics in realistic photosynthetic
antenna are still lacking, because the typical size of an antenna
($>100$ chromophores) and the warm and wet environment make the accurate
description of coherent EET in these systems a formidable task. Methods
for simulating coherent EET dynamics in photosynthetic systems have
been put forward (see \cite{OlayaCastro:2011hw} for a comprehensive
review). For example, Förster theory directly provides EET rates,
however it neglects quantum coherence effects that have been shown
to be critical in photosynthetic systems \cite{Cheng:2009p75665}.
In contrast, Redfield theory describes coherent EET dynamics, but
it is valid only when the system-bath interactions are sufficiently
weak \cite{Cao:2000up,Ishizaki:2009p75286}. In order to bridge the
gap between these two methods, recent advances in theoretical methods
have focused on new approaches valid in the intermediate regime. For
example, the small-polaron master equation approach \cite{Jang:2008ef,Jang:2011ee,Kolli:2011ki,Chang:2012il}
should provide a reasonable description of the EET dynamics in the
intermediate regime, and the hierarchical equation of motion (HEOM)
actually yields a numerically exact means to calculate EET dynamics
\cite{Tanimura:2006p3161,Jin:2008p60247,Ishizaki:2009p75287}. 

Nevertheless, it is still difficult to apply these theoretical methods
to a realistic photosynthetic antenna, which often exhibits $>100$
chromophores and static disorder, due to the formidable computational
resources required. Moreover, to understand the fundamentals of light-matter
interactions in photosynthetic light harvesting and realize laser
control of energy flow in photosynthetic systems, explicit treatment
of photo-excitation processes would be necessary. Furthermore, it
was suggested recently that the photoexcitation condition determined
by the nature of the photon source could strongly affect the appearance
of EET dynamics in photosynthetic complexes \cite{Jiang:1991kc,Hoki:2011kc,Mancal:2010kc,Brumer:2012ib,Fassioli:2012gd}.
As a result, the interplay between the photoexcitation process and
ultrafast EET dynamics may be highly nontrivial, and it would be important
to explicitly consider system-field interactions in a theoretical
description of light harvesting. Therefore, an accurate yet numerically
efficient method for simulating quantum dynamics in photosynthetic
systems in the presence of coherent laser fields is highly desirable. 

In this work, we combine a coherent modified Redfield theory (CMRT)
\cite{YuHsienHwangFu:2012we} and a non-Markovian quantum jump (NMQJ)
method \cite{Piilo:2008di,Piilo:2009p89261,Rebentrost:2009kv} to
develop an efficient approach for coherent EET dynamics in photosynthetic
systems under the influence of light-matter interactions. The modified-Redfield
theory \cite{Zhang:1998p3588,Yang:2002p3160} yields reliable population
dynamics of excitonic systems and has been successfully applied to
describe spectra and population dynamics in many photosynthetic complexes.\cite{Novoderezhkin:2006p516,Novoderezhkin:2010fb,Novoderezhkin:2010p82122}
As a generalization of the modified-Redfield theory, the CMRT approach
provides a complete description of quantum dynamics of the system's
density matrix. Furthermore, we adopt a quantum jump method to achieve
efficient propagation of the CMRT equations of motion. Using quantum
jumps or quantum trajectories to unravel a set of equations of motion
for the density matrix into a stochastic Schr\"{o}dinger equation
has proven to be a powerful tool for the simulation of quantum dynamics
in an open quantum system \cite{Plenio:1998ul,Diosi:1998um,Strunz:1999tj,Brun:2002p93736,Piilo:2008di,Piilo:2009p89261}.
In particular, Piilo \emph{et al.} have developed a non-Markovian
quantum jump (NMQJ) method that can be implemented for simulating
non-Markovian dynamics in multilevel systems \cite{Piilo:2008di,Piilo:2009p89261}.
To make use of the NMQJ method, which has been shown to be efficient
for simulating quantum dynamics in EET \cite{Rebentrost:2009kv,Ai:2013wl},
we recast the CMRT equation into a generalized Lindblad form \cite{Breuer:2002wp}.
In addition, by transforming to a rotating frame, we extend CMRT to
describe a time-dependent Hamiltonian, which is necessary to describe
systems in the presence of laser fields. To examine the validity of
the combined CMRT-NMQJ approach, we simulate the EET dynamics in the
FMO complex. Moreover, we investigate a simple dimer system under
the influence of laser pulses. We aim to shed light on how the photoexcitation
process affects the coherence of EET dynamics in light harvesting. 

This paper is organized as follows. In the next section, we briefly
describe CMRT for EET in a dimer system with time-independent Hamiltonian.
Then, we extend the CMRT to treat time-dependent case by including
interactions with a laser pulse applied to the dimer. Furthermore,
in order to make the CMRT suitable for time-propagation by the NMQJ
method, we show that the CMRT quantum master equation can be rewritten
in a generalized Lindblad form. In section 3, we detail the implementation
of the NMQJ method to solve the CMRT master equation. To verify the
applicability of the new method, in section 4, we apply it to simulate
the EET dynamics in FMO and compare the results to those from numerically
exact simulations. Furthermore, in section 5, we calculate the dynamics
of site populations and entanglement in a dimer system after it has
been excited by a laser pulse to investigate the interplay between
laser excitation and ultrafast EET in the system. Finally, the main
results of this work are summarized in section 6.

\section{Coherent Modified Redfield Theory}

\label{sec:CMRT}

For the sake of self-consistency, we briefly describe the CMRT method
in this section. Note that the focus of this work is the combined
CMRT-NMQJ approach. The details in the derivation and validity of
the CMRT approach is outside of the scope of this paper and will be
published elsewhere \cite{YuHsienHwangFu:2012we}.

\subsection{Time-Independent Hamiltonian}

\begin{figure}
\centering\includegraphics[bb=30bp 800bp 480bp 1040bp,clip,width=7cm]{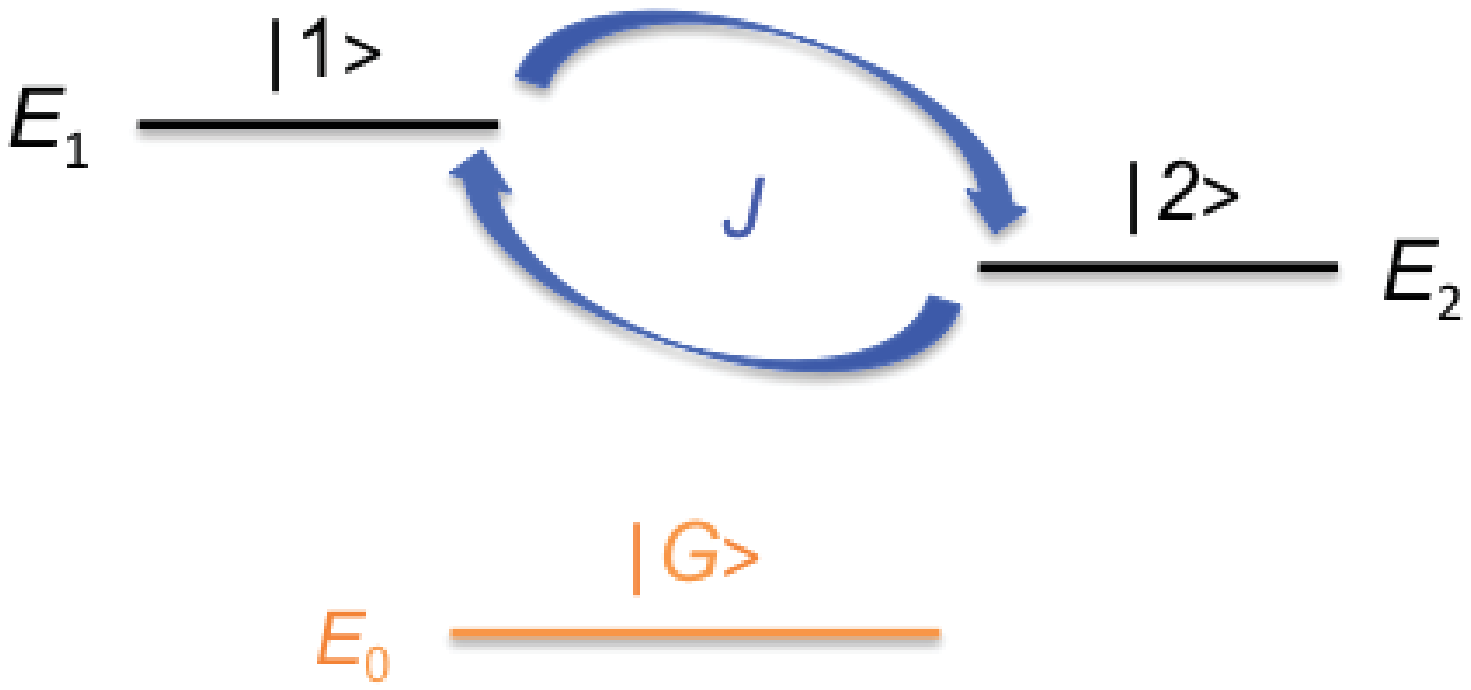}\caption{Energy diagram of a dimer for a time-independent case. The site energy
for the single-excitation state $\left\vert n\right\rangle $ is $E_{n}$
and the energy of the ground state $\left\vert G\right\rangle $ is
$E_{0}$. The electronic coupling $J$ between the single-excitation
states is also considered.\label{fig:Energy-Diagram-Free}}
\end{figure}

Before proceeding to the case with a time-dependent Hamiltonian, we
briefly review the CMRT approach for a time-independent Hamiltonian.
The CMRT approach is a generalization of the original modified Redfield
theory \cite{Zhang:1998p3588,Yang:2002p3160} to treat coherent evolution
of the full density matrix of an excitonic system. For simplicity,
we investigate the quantum dynamics in a dimer system (figure \ref{fig:Energy-Diagram-Free})
described by the following excitonic Hamiltonian:

\begin{equation}
H_{S}^{(1)}=\sum_{n=1}^{2}E_{n}\left\vert n\right\rangle \left\langle n\right\vert +J\left(\left\vert 1\right\rangle \left\langle 2\right\vert +\left\vert 2\right\rangle \left\langle 1\right\vert \right)+E_{0}\left\vert G\right\rangle \left\langle G\right\vert ,
\end{equation}
where $\left\vert n\right\rangle $ is the product state of the excited
state on the $n$th site and the ground state on the other site, i.e.,
$\left|1\right\rangle =\left|e\right\rangle _{1}\left|g\right\rangle _{2}$,
$\left|2\right\rangle =\left|g\right\rangle _{1}\left|e\right\rangle _{2}$,
$E_{n}$ is the corresponding site energy, $J$ is the electronic
coupling between these two sites. Here, when there is no pulse applied
to the system, the ground state $\left\vert G\right\rangle =\left|g\right\rangle _{1}\left|g\right\rangle _{2}$
with energy $E_{0}$ is decoupled to the single-excitation states
$\left\vert n\right\rangle $. Note that although the Hamiltonian
for a dimer system is considered in this work, the results can be
straightforwardly generalized to treat multi-site systems, e.g. the
FMO with seven sites as shown in section \ref{sec:FMO}.

Furthermore, the environment is modeled as a collection of harmonic
oscillators described by the Hamiltonian 
\begin{equation}
H_{B}=\sum_{n=1}^{2}\sum_{q}\left[\frac{p_{nq}^{2}}{2m_{q}}+\frac{1}{2}m_{q}\omega_{q}^{2}x_{nq}^{2}\right],
\end{equation}
where $p_{nq}$ and $x_{nq}$ are respectively the momentum and position
for $q$th harmonic oscillator on the $n$th site with mass $m_{q}$
and frequency $\omega_{q}$. Without loss of generality, we have assumed
an independent bath for each site. Nevertheless, the CMRT can deal
with a general system with correlated baths, whose effects on EET
dynamics have been investigated in detail previously \cite{Wu:2010bg,Chang:2012il,Wu:2012ch}. 

To describe exciton-bath interactions, we assume linear system-bath
couplings described by
\begin{eqnarray}
H_{SB} & = & \sum_{n=1}^{2}u_{n}\left\vert n\right\rangle \left\langle n\right\vert ,\label{eq:HSB}
\end{eqnarray}
where the collective bath-dependent coupling is 
\begin{eqnarray}
u_{n} & = & \sum_{q}m_{q}\omega_{q}^{2}x_{nq}^{(0)}x_{nq},
\end{eqnarray}
where $x_{nq}^{(0)}$ denotes the displacement of the $q$th harmonic
mode when the $n$th site is excited.For the electronic part, we diagonalize
the system Hamiltonian in the exciton basis \cite{Cheng:2009p75665}
\begin{equation}
H_{S}^{(1)}=\sum_{k=0}^{2}\varepsilon_{k}^{\prime}\left\vert \varepsilon_{k}^{(1)}\right\rangle \left\langle \varepsilon_{k}^{(1)}\right\vert ,
\end{equation}
where each exciton basis state is a superposition of site-localized
excitations
\begin{equation}
\left\vert \varepsilon_{k}^{(1)}\right\rangle =\sum_{n=0}^{2}C_{kn}^{(1)}\left\vert n\right\rangle ,
\end{equation}
with exciton energy $\varepsilon_{k}^{\prime}$. Since there is no
interaction between the ground state and the single-excitation states,
we have 
\begin{equation}
\left\vert \varepsilon_{0}^{(1)}\right\rangle =\left\vert 0\right\rangle =\left\vert G\right\rangle 
\end{equation}
with
\begin{equation}
\varepsilon_{0}^{\prime}=E_{0}.
\end{equation}
In other words, 
\begin{equation}
C_{0n}^{(1)}=\delta_{n0}.\label{eq:C0n}
\end{equation}
Hereafter, for the sake of simplicity, we shall label $|G\rangle$
as $|0\rangle$.

Transforming $H_{SB}$ to the exciton basis, the system-bath interactions
contain both diagonal and off-diagonal terms. Following the essence
of the modified Redfield theory \cite{Zhang:1998p3588}, we include
the diagonal term of $H_{SB}$ into the zeroth-order Hamiltonian in
the exciton basis to be treated nonperturbatively, and we only treat
the off-diagonal term of $H_{SB}$ in the exciton basis as the perturbation.
In the case, the unperturbed and perturbation Hamiltonians read respectively
\begin{eqnarray}
H_{0}^{(1)} & = & \sum_{k=0}^{2}\left[\varepsilon_{k}^{\prime}+H_{B}+\sum_{n=1}^{2}u_{n}a_{kk}^{(1)}(n)\right]\left\vert \varepsilon_{k}^{(1)}\right\rangle \left\langle \varepsilon_{k}^{(1)}\right\vert ,\\
V^{(1)} & = & \sum_{k=0}^{2}\sum_{k^{\prime}=0\left(\neq k^{\prime}\right)}^{2}\sum_{n=1}^{2}u_{n}a_{kk^{\prime}}^{(1)}(n)\left\vert \varepsilon_{k}^{(1)}\right\rangle \left\langle \varepsilon_{k^{\prime}}^{(1)}\right\vert ,
\end{eqnarray}
where 
\begin{eqnarray}
a_{kk^{\prime}}^{(1)}(n) & = & C_{kn}^{(1)*}C_{k^{\prime}n}^{(1)}.
\end{eqnarray}
Notice that $a_{kk^{\prime}}^{(1)}(n)=0$ as long as $k=0$ or $k^{\prime}=0$
as a result of (\ref{eq:C0n}). We then apply the second-order cumulant
expansion approach using $V^{(1)}$ as the perturbation followed by
secular approximation to derive a quantum master equation for the
reduced density matrix of the electronic system\cite{Zhang:1998p3588,Mukamel:1999us,YuHsienHwangFu:2012we}:
\begin{eqnarray}
\fl\partial_{t}\rho=-\rmi\sum_{k,k^{\prime}}\left(\varepsilon_{k}^{(1)}-\varepsilon_{k^{\prime}}^{(1)}\right)\rho_{kk^{\prime}}\left\vert \varepsilon_{k}^{(1)}\right\rangle \left\langle \varepsilon_{k^{\prime}}^{(1)}\right\vert \nonumber \\
-\frac{1}{2}\sum_{k,k^{\prime}}\left[R_{k^{\prime}k}^{(1)\textrm{dis}}(t)\rho_{kk}\left\vert \varepsilon_{k}^{(1)}\right\rangle \left\langle \varepsilon_{k}^{(1)}\right\vert -R_{kk^{\prime}}^{(1)\textrm{dis}}(t)\rho_{k^{\prime}k^{\prime}}\left\vert \varepsilon_{k^{\prime}}^{(1)}\right\rangle \left\langle \varepsilon_{k^{\prime}}^{(1)}\right\vert \right]\nonumber \\
-\sum_{k,k^{\prime}}\left\{ R_{kk^{\prime}}^{(1)\textrm{pd}}(t)+\frac{1}{2}\sum_{k^{\prime\prime}}\left[R_{k^{\prime\prime}k}^{(1)\textrm{dis}}(t)+R_{k^{\prime\prime}k^{\prime}}^{(1)\textrm{dis}}(t)\right]\right\} \rho_{kk^{\prime}}\left\vert \varepsilon_{k}^{(1)}\right\rangle \left\langle \varepsilon_{k^{\prime}}^{(1)}\right\vert ,\label{eq:OriEq}
\end{eqnarray}
where 
\begin{eqnarray}
\varepsilon_{k}^{(1)} & = & \varepsilon_{k}^{\prime}-\Lambda_{k}^{(1)},\\
\Lambda_{k}^{(1)} & = & \sum_{n=1}^{2}\left[a_{kk}^{(1)}(n)\right]^{2}\lambda_{n},
\end{eqnarray}
and the reorganization energy is
\begin{equation}
\lambda_{n}=\sum_{q}\frac{1}{2}m_{q}\omega_{q}^{2}x_{nq}^{(0)2}.
\end{equation}
Here $\varepsilon_{k}^{(1)}$ can be viewed as the eigen value of
the electronic Hamiltonian shifted by bath reorganization.The CMRT
master equation (\ref{eq:OriEq}) describes the coherent dynamics
driven by the exciton Hamiltonian, the population transfer dynamics,
and dephasing dynamics including pure-dephasing induced by diagonal
fluctuations and population-transfer induced dephasing. The dissipation
rate $R_{kk^{\prime}}^{(1)\textrm{dis}}(t)$ is defined as
\begin{eqnarray}
\fl R_{kk^{\prime}}^{(1)\textrm{dis}}\left(t\right)=2\textrm{Re}\int_{0}^{t}d\tau\rme^{\rmi\left(\varepsilon_{k^{\prime}}^{(1)}-\varepsilon_{k}^{(1)}\right)\tau}\rme^{-\rmi\left(\Lambda_{k}^{(1)}+\Lambda_{k^{\prime}}^{(1)}-2\lambda_{kk,k^{\prime}k^{\prime}}^{(1)}\right)\tau}\rme^{-g_{kkkk}^{(1)}(\tau)-g_{k^{\prime}k^{\prime}k^{\prime}k^{\prime}}^{(1)}(\tau)+2g_{kk,k^{\prime}k^{\prime}}^{(1)}(\tau)}\nonumber \\
\times\{\ddot{g}_{k^{\prime}k,kk^{\prime}}^{(1)}(\tau)-[\dot{g}_{k^{\prime}k,kk}^{(1)}(\tau)-\dot{g}_{k^{\prime}k,k^{\prime}k^{\prime}}^{(1)}(\tau)-2\rmi\lambda_{k^{\prime}k,k^{\prime}k^{\prime}}^{(1)}]\nonumber \\
\times[\dot{g}_{kk^{\prime},kk}^{(1)}(\tau)-\dot{g}_{kk^{\prime},k^{\prime}k^{\prime}}^{(1)}(\tau)-2\rmi\lambda_{kk^{\prime},k^{\prime}k^{\prime}}^{(1)}]\},\label{eq:disR}
\end{eqnarray}
where $g_{k_{1}k_{2}k_{3}k_{4}}^{(1)}(t)$ is the line-broadening
function evaluated from the spectral density of the system-bath couplings
\cite{Mukamel:1999us}:
\begin{eqnarray}
g_{k_{1}k_{2}k_{3}k_{4}}^{(1)}(t) & = & \sum_{n}a_{k_{1}k_{2}}^{(1)}(n)a_{k_{3}k_{4}}^{(1)}(n)g_{n}(t),\\
\lambda_{k_{1}k_{2}k_{3}k_{4}}^{(1)} & = & \sum_{n}a_{k_{1}k_{2}}^{(1)}(n)a_{k_{3}k_{4}}^{(1)}(n)\lambda_{n},\\
g_{n}(t) & = & \int d\omega\frac{J_{n}(\omega)}{\omega^{2}}\left[\left(1-\cos\omega t\right)\coth\left(\frac{\beta\omega}{2}\right)+\rmi\left(\sin\omega t-\omega t\right)\right]
\end{eqnarray}
 In addition, the pure-dephasing rate $R_{kk^{\prime}}^{(1)\textrm{pd}}(t)$
is given by
\begin{eqnarray}
R_{kk^{\prime}}^{(1)\textrm{pd}}(t) & = & \sum_{n=1}^{2}\left[a_{kk}^{(1)}(n)-a_{k^{\prime}k^{\prime}}^{(1)}(n)\right]^{2}\textrm{Re}\left[\dot{g}_{n}(t)\right].\label{eq:pdR}
\end{eqnarray}

The CMRT equation (\ref{eq:OriEq}) can be considered as a generalization
of the original modified Redfield theory to treat full coherent dynamics
of the excitonic system. By treating the diagonal system-bath coupling
in the exciton basis nonperturbatively, the CMRT considers multi-phonon
effects and generally interpolates between the strong system-bath
coupling and the weak system-bath coupling limits, in contrast to
the Redfield theory \cite{Yang:2002p3160,Novoderezhkin:2010p82122}.
Despite this, our method still shows its promising validity as compared
to the numerically exact HEOM approach shown in section \ref{sec:FMO}.

Functions required to propagate the CMRT master equation can be readily
calculated if the system Hamiltonian $H_{S}$ and the spectral densities
$J_{n}(\omega)$ are provided. Generally speaking, the spectral densities
used in theoretical investigations of photosynthetic systems are obtained
by fitting to available experimental data, e.g., linear absorption
lineshape or nonlinear spectra \cite{Cheng:2009p75665,Renger:el}.
For self-consistency, the same theoretical method should be used in
the fitting process of the parameters and dynamical simulations. Therefore,
it is important that a dynamical method can also describe spectroscopy
within the same framework. For example, the absorption lineshape defined
as the Fourier transform of the dipole-dipole correlation function
can be calculated, i.e.,
\begin{equation}
I(\omega)\equiv\textrm{Re}\left[\int_{0}^{\infty}dt\chi(t)\rme^{\rmi\omega t}\right],
\end{equation}
where the dipole-dipole correlation function is 
\begin{equation}
\chi(t)\equiv\textrm{Tr}\left[\mu(t)\mu(0)\right].
\end{equation}
Within the framework of the CMRT under the same second-order approximation,
the dipole-dipole correlation function reads \cite{Ohta:2001bh,Novoderezhkin:2006p516}
\begin{equation}
\chi\simeq\sum_{k}\left|\mu_{kG}\right|^{2}\exp\left[\rmi\left(\varepsilon_{0}^{(1)}-\varepsilon_{k}^{(1)}\right)t-g_{kkkk}^{(1)}(t)-\frac{1}{2}\sum_{k^{\prime}\neq k}R_{k^{\prime}k}^{(1)\textrm{dis}}\left(\infty\right)t\right].\label{eq:chi}
\end{equation}
It corresponds to several peaks around the level spacings between
the exciton states and the ground state, i.e., $\varepsilon_{k}^{(1)}-\varepsilon_{0}^{(1)}$.
The second term in the exponential $g_{kkkk}^{(1)}(t)$ is the exchange-narrowed
line-shape function \cite{Ohta:2001bh}, which is proportional to
the participation ratio $\sum_{n}\left|C_{kn}^{(1)}\right|^{4}$.
Generally speaking, for a delocalized state $\sum_{n}\left|C_{kn}^{(1)}\right|^{4}$
is much smaller than that for a completely-localized state. Therefore,
the widths of the peaks in the absorption line shape are significantly
narrowed due to exciton delocalization. The third term in the exponential
is the relaxation-induced broadening. It originates from the population
relaxation within the single-excitation subspace when the molecule
is excited by light, and has been demonstrated to be important in
the spectra of molecular assemblies \cite{Novoderezhkin:2006p516}.

\subsection{Coherent Modified Redfield Theory in the Presence of Laser Fields}

\label{sec:CMRTCP}

\begin{figure}
\centering\includegraphics[bb=30bp 700bp 480bp 1040bp,clip,width=7cm]{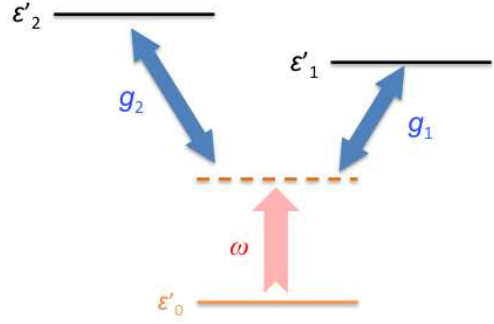}\caption{Schematic diagram of a dimer with a laser pulse applied. In the rotating
frame, the ground state $\left\vert \varepsilon_{0}^{(1)}\right\rangle $
is lifted to $\varepsilon_{0}^{\prime}+\omega$. And there are laser-induced
transitions between the excited states $\left\vert \varepsilon_{k}^{(1)}\right\rangle $
and $\left\vert \varepsilon_{0}^{(1)}\right\rangle $ with coupling
strength $g_{k}$.\label{fig:Energy-Diagram-Pulsed}}
\end{figure}

To treat system-field interactions in the framework of the CMRT, we
need to extend the CMRT equations of motion to include influences
of laser fields. As shown in figure \ref{fig:Energy-Diagram-Pulsed},
during the time-evolution, we consider a dimer system interacting
with a monochromatic pulse whose frequency is $\omega$, thus inducing
couplings between the ground state $\left\vert \varepsilon_{0}^{(1)}\right\rangle $
and delocalized exciton states $\left\vert \varepsilon_{k}^{(1)}\right\rangle $
($k=1,2$). In this case, the electronic Hamiltonian reads 
\begin{equation}
H_{S}^{(2)}=\sum_{k=0}^{2}\varepsilon_{k}^{\prime}\left\vert \varepsilon_{k}^{(1)}\right\rangle \left\langle \varepsilon_{k}^{(1)}\right\vert +2\cos\omega t\sum_{k=1}^{2}g_{k}\left(\left\vert \varepsilon_{k}^{(1)}\right\rangle \left\langle \varepsilon_{0}^{(1)}\right\vert +\textrm{h.c.}\right),
\end{equation}
where $2g_{k}$ is the laser-induced coupling strength between the
ground state $\left\vert \varepsilon_{0}^{(1)}\right\rangle $ and
the $k$th delocalized exciton state $\left\vert \varepsilon_{k}^{(1)}\right\rangle $.
Transformed to a rotating frame with
\[
U=\exp\left[\rmi\omega t\left\vert \varepsilon_{0}^{(1)}\right\rangle \left\langle \varepsilon_{0}^{(1)}\right\vert \right],
\]
the effective Hamiltonian $H_{\mathrm{eff}}^{(2)}=U^{\dagger}H_{S}^{(2)}U+\rmi\dot{U}^{\dagger}U$
of the electronic part becomes
\begin{eqnarray}
\fl H_{\mathrm{eff}}^{(2)}\simeq\sum_{k=1}^{2}\varepsilon_{k}^{\prime}\left\vert \varepsilon_{k}^{(1)}\right\rangle \left\langle \varepsilon_{k}^{(1)}\right\vert +\left(\varepsilon_{0}^{\prime}+\omega\right)\left\vert \varepsilon_{0}^{(1)}\right\rangle \left\langle \varepsilon_{0}^{(1)}\right\vert +\sum_{k=1}^{2}g_{k}\left(\left\vert \varepsilon_{k}^{(1)}\right\rangle \left\langle \varepsilon_{0}^{(1)}\right\vert +\textrm{h.c.}\right),
\end{eqnarray}
where we have applied the rotating-wave approximation and dropped
the fast-oscillating terms with factors $\exp\left(\pm i2\omega t\right)$.
The above Hamiltonian can be diagonalized as
\begin{equation}
H_{\mathrm{eff}}^{(2)}=\sum_{k=0}^{2}\varepsilon_{k}^{\prime\prime}\left\vert \varepsilon_{k}^{(2)}\right\rangle \left\langle \varepsilon_{k}^{(2)}\right\vert ,\label{eq:Heff}
\end{equation}
where 
\begin{equation}
\left\vert \varepsilon_{k}^{(2)}\right\rangle =\sum_{k^{\prime}=0}^{2}C_{kk^{\prime}}^{(2)}\left\vert \varepsilon_{k^{\prime}}^{(1)}\right\rangle =\sum_{n=0}^{2}\sum_{k^{\prime}=0}^{2}C_{kk^{\prime}}^{(2)}C_{k^{\prime}n}^{(1)}\left\vert n\right\rangle 
\end{equation}
is the electronic eigen state with eigen energy $\varepsilon_{k}^{\prime\prime}$.

In the basis of $\left\{ \left\vert \varepsilon_{k}^{(2)}\right\rangle \right\} $
in the rotating frame, the unperturbed and perturbation Hamiltonians
read respectively 
\begin{eqnarray}
H_{0}^{(2)} & = & \sum_{k=0}^{2}\left[\varepsilon_{k}^{\prime\prime}+H_{B}+\sum_{n=1}^{2}u_{n}a_{kk}^{(2)}(n)\right]\left\vert \varepsilon_{k}^{(2)}\right\rangle \left\langle \varepsilon_{k}^{(2)}\right\vert ,\nonumber \\
V^{(2)} & = & \sum_{k=0}^{2}\sum_{k^{\prime}=0\left(\neq k\right)}^{2}\sum_{n=1}^{2}u_{n}a_{kk^{\prime}}^{(2)}(n)\left\vert \varepsilon_{k}^{(2)}\right\rangle \left\langle \varepsilon_{k^{\prime}}^{(2)}\right\vert ,
\end{eqnarray}
where 
\begin{eqnarray}
a_{kk^{\prime}}^{(2)}(n) & = & \sum_{k_{1}=0}^{2}C_{kk_{1}}^{(2)*}C_{k_{1}n}^{(1)*}\sum_{k_{2}=0}^{2}C_{k^{\prime}k_{2}}^{(2)}C_{k_{2}n}^{(1)}.
\end{eqnarray}

Following the CMRT approach, we could obtain the master equation 
\begin{eqnarray}
\fl\partial_{t}\rho^{\mathrm{T}}=-\rmi\sum_{k,k^{\prime}}\left(\varepsilon_{k}^{(2)}-\varepsilon_{k^{\prime}}^{(2)}\right)\rho_{kk^{\prime}}^{\mathrm{T}}\left\vert \varepsilon_{k}^{(2)}\right\rangle \left\langle \varepsilon_{k^{\prime}}^{(2)}\right\vert \nonumber \\
-\frac{1}{2}\sum_{k,k^{\prime}}\left[R_{k^{\prime}k}^{(2)\textrm{dis}}(t)\rho_{kk}^{\mathrm{T}}\left\vert \varepsilon_{k}^{(2)}\right\rangle \left\langle \varepsilon_{k}^{(2)}\right\vert -R_{kk^{\prime}}^{(2)\textrm{dis}}(t)\rho_{k^{\prime}k^{\prime}}^{\mathrm{T}}\left\vert \varepsilon_{k^{\prime}}^{(2)}\right\rangle \left\langle \varepsilon_{k^{\prime}}^{(2)}\right\vert \right]\nonumber \\
-\sum_{k,k^{\prime}}\left\{ R_{kk^{\prime}}^{(2)\textrm{pd}}(t)+\frac{1}{2}\sum_{k^{\prime\prime}}\left[R_{k^{\prime\prime}k}^{(2)\textrm{dis}}(t)+R_{k^{\prime\prime}k^{\prime}}^{(2)\textrm{dis}}(t)\right]\right\} \rho_{kk^{\prime}}^{\mathrm{T}}\left\vert \varepsilon_{k}^{(2)}\right\rangle \left\langle \varepsilon_{k^{\prime}}^{(2)}\right\vert ,\label{eq:OriEq-1}
\end{eqnarray}
where $\rho^{\mathrm{T}}(t)=U^{\dagger}\rho(t)U$ is the density matrix
in the rotating frame, 
\begin{eqnarray}
\varepsilon_{k}^{(2)} & = & \varepsilon_{k}^{\prime\prime}-\Lambda_{k}^{(2)},\\
\Lambda_{k}^{(2)} & = & \sum_{n=1}^{2}\left[a_{kk}^{(2)}(n)\right]^{2}\lambda_{n}.
\end{eqnarray}
Here $\varepsilon_{k}^{(2)}$ can be viewed as the eigen value of
the electronic Hamiltonian 
\begin{eqnarray}
H_{e}^{(2)} & = & \sum_{k}\varepsilon_{k}^{(2)}\left\vert \varepsilon_{k}^{(2)}\right\rangle \left\langle \varepsilon_{k^{\prime}}^{(2)}\right\vert .
\end{eqnarray}
In (\ref{eq:OriEq-1}), the dissipation rate $R_{kk^{\prime}}^{(2)\textrm{dis}}\left(t\right)$
and pure-dephasing rate $R_{kk^{\prime}}^{(2)\textrm{pd}}(t)$ can
be obtained following (\ref{eq:disR}) and (\ref{eq:pdR}) by substituting
$\varepsilon_{k}^{(1)}$, $\Lambda_{k}^{(1)}$, $g_{k_{1}k_{2}k_{3}k_{4}}^{(1)}(t)$,
$\lambda_{k_{1}k_{2}k_{3}k_{4}}^{(1)}$ respectively with $\varepsilon_{k}^{(2)}$,
$\Lambda_{k}^{(2)}$, 
\begin{eqnarray}
g_{k_{1}k_{2}k_{3}k_{4}}^{(2)}(t) & = & \sum_{n}a_{k_{1}k_{2}}^{(2)}(n)a_{k_{3}k_{4}}^{(2)}(n)g_{n}(t),\\
\lambda_{k_{1}k_{2}k_{3}k_{4}}^{(2)} & = & \sum_{n}a_{k_{1}k_{2}}^{(2)}(n)a_{k_{3}k_{4}}^{(2)}(n)\lambda_{n}.
\end{eqnarray}

Thus, the influence of the laser field can be considered as modifying
the electronic part of the Hamiltonian by pulse induced couplings
between ground state and single-excitation states in the rotating
frame. Notice that due to the transformation, the density matrix in
the original frame $\rho(t)$ is transformed to $\rho^{\mathrm{T}}(t)=U^{\dagger}\rho(t)U$.
Hence, after solving the master equation, the obtained density matrix
should be transformed back to the original frame as $\rho(t)=U\rho^{\mathrm{T}}(t)U^{\dagger}$.
We further remark that (\ref{eq:OriEq-1}) enables us to simulate
field-induced and dissipative dynamics for an excitonic system due
to a step-function pulse. However, this is not a limitation because
of the time-local nature of the equations of motion. Within any propagation
scheme based on discrete time steps, this formalism can be easily
generalized to arbitrary pulse shapes as discussed in \ref{sec:appGAPS}.

\subsection{Lindblad-form Coherent Modified Redfield Theory}

\label{sec:LCMRT}

We have obtained the CMRT master equation that governs the dynamics
of an excitonic system in the presence of laser fields. For a system
with $M$ chromophores, the CMRT master equation corresponds to a
set of $M^{2}$ ordinary differential equations. As a consequence,
it would be a formidable task to propagate the dynamics if the system
has hundreds of sites, just as in a typical photosynthetic complex,
e.g., 96 chlorophylls in PSI. We can apply the NMQJ method \cite{Piilo:2008di,Piilo:2009p89261}
to reduce the computational complexity to the order of $M$ to achieve
efficient simulation of the CMRT dynamics. The NMQJ method propagates
a set of equations of motion in the generalized Lindblad form \cite{Breuer:2007tx}:
\begin{eqnarray}
\fl\partial_{t}\rho & = & -\rmi[H_{e}(t),\rho]-\frac{1}{2}\sum_{k,k^{\prime}}R_{kk^{\prime}}(t)\left[\left\{ A_{kk^{\prime}}^{\dagger}(t)A_{kk^{\prime}}(t),\rho\right\} -2A_{kk^{\prime}}(t)\rho A_{kk^{\prime}}^{\dagger}(t)\right].\label{eq:LCMRT}
\end{eqnarray}
Thus, in order to solve the master equation (cf. (\ref{eq:OriEq})
and (\ref{eq:OriEq-1})) by means of the NMQJ method, we should rewrite
them in the generalized Lindblad form. To this end, we define the
following jump operators
\begin{equation}
A_{kk^{\prime}}(t)=\left\vert \varepsilon_{k}(t)\right\rangle \left\langle \varepsilon_{k^{\prime}}(t)\right\vert ,
\end{equation}
where $\left\vert \varepsilon_{k}(t)\right\rangle $ is $k$th eigen
state of the system Hamiltonian $H_{e}(t)$, and the matrix elements
of the population transfer and dephasing rates are defined respectively
as\begin{eqnarray} R_{kk^{\prime}} & \equiv & \cases{R_{kk^{\prime}}^{\textrm{dis}}, &\textrm{for} $ k\neq k^{\prime}$,\\ \Gamma_{k}, & \textrm{for} $k=k^{\prime}$,}\label{eq:RateLCMRT} \end{eqnarray}where
$\Gamma_{k}$s are determined by the pure-dephasing rates in the CMRT
master equation: 
\begin{equation}
\Gamma=B^{-1}A,
\end{equation}
where the matrix elements of $A$ and $B$ are respectively
\begin{equation}
A_{a}=\sum_{k=a+1}^{M}R_{ak}^{\textrm{pd}}+\sum_{k=1}^{M-1}R_{ka}^{\textrm{pd}},
\end{equation}
\begin{equation} B_{jk}=\cases{ \frac{1}{2}, &\textrm{for} $k<j$,\\ \frac{1}{2}(2M-j), &\textrm{for} $k=j$,\\ 1, &\textrm{for} $j<k<M$,\\ \frac{1}{2}, &\textrm{for} otherwise.} \end{equation}Note
that since there are $M(M-1)/$2 independent pure-dephasing rates
$R_{kk^{\prime}}^{\textrm{pd}}$ in the CMRT master equation and only
$M$ Lindblad-form dephasing rates $\Gamma_{k}$, the problem is overdetermined.
We found that a least-square fit of $\Gamma_{k}$ to $R_{kk^{\prime}}^{\textrm{pd}}$
preserves the CMRT dynamics extremely well and is also numerically
straightforward to implement. The details of the numerical determination
of Lindblad-form dephasing rates are presented in \ref{sec:appELF}.

\section{Non-Markovian Quantum Jump Method}

\label{sec:NMQTM}

We apply the non-Markovian quantum jump (NMQJ) approach proposed by
Piilo \emph{et al.} to simulate the CMRT equation of motion for multilevel
systems \cite{Piilo:2008di,Piilo:2009p89261}. The original implementation
of the NMQJ approach only considers a time-independent Hamiltonian.
In order to apply the NMQJ method to simulate CMRT dynamics under
the influence of laser fields, here we briefly describe the NMQJ method
and show how to explicitly implement the method for the time-dependent
case to propagate the quantum dynamics of an $M$-level system under
the influence of external time-dependent fields.

In our theoretical investigation, we consider the system to be initially
in the ground state $\left\vert \psi_{M}(0)\right\rangle =\left\vert G\right\rangle $
and thus the initial density matrix is
\begin{equation}
\rho(0)=\left\vert \psi_{M}(0)\right\rangle \left\langle \psi_{M}(0)\right\vert .
\end{equation}
Then, the laser is turned on and applied to the system for a duration
$\tau$ starting at $t=0$. According to the NMQJ method, for an $M$-level
system, there are $M+1$ possible states for the system to propagate
in, i.e., $\left\vert \varepsilon_{k}^{(2)}\right\rangle $ for $k=0,1,\cdots M-1$,
in addition to the deterministic evolution
\begin{equation}
\left\vert \psi_{M}^{\mathrm{T}}(t+\delta t)\right\rangle =e^{-\rmi H(t)\delta t}\left\vert \psi_{M}^{\mathrm{T}}(t)\right\rangle \left\Vert e^{-\rmi H(t)\delta t}\left\vert \psi_{M}^{\mathrm{T}}(t)\right\rangle \right\Vert ^{-1},
\end{equation}
where the non-Hermitian Hamiltonian is
\begin{equation}
H(t)=H_{e}(t)-\frac{\rmi}{2}\sum_{k,k^{\prime}}R_{kk^{\prime}}(t)A_{kk^{\prime}}^{\dagger}(t)A_{kk^{\prime}}(t).
\end{equation}
If $N$ ensemble members are considered in the simulation, the NMQJ
method describes the number of ensemble members in each of the individual
states at time $t$, $N_{\alpha}^{(\mathrm{I})}(t)$ ($\alpha=0,1,\cdots M$),
with the initial value given as\begin{equation}N_{\alpha}^{(\mathrm{I})}(0)= \cases{0,&\textrm{for} $\alpha=0,1,\cdots M-1$,\\N,&\textrm{for} $\alpha=M$.}\end{equation}
During the pulse duration, the Lindblad-form master equation (\ref{eq:OriEq-1})
is solved by the NMQJ method to obtain the distribution of $N_{\alpha}^{(\mathrm{I})}(t)$.
Therefore, the density matrix in the rotating frame is straightforwardly
given by
\begin{equation}
\rho^{\mathrm{T}}(t)=\frac{1}{N}\left[\sum_{k=0}^{M-1}N_{k}^{(\mathrm{I})}(t)\left\vert \varepsilon_{k}^{(2)}\right\rangle \left\langle \varepsilon_{k}^{(2)}\right\vert +N_{M}^{(\mathrm{I})}(t)\left\vert \psi_{M}^{\mathrm{T}}(t)\right\rangle \left\langle \psi_{M}^{\mathrm{T}}(t)\right\vert \right].
\end{equation}
At the end of the pulse, we transform the density matrix back to the
original frame by $\rho(\tau)=U(\tau)\rho^{\mathrm{T}}(\tau)U^{\dagger}(\tau)$
to yield 
\begin{equation}
\left\vert \psi_{M}(\tau)\right\rangle =U(\tau)\left\vert \psi_{M}^{\mathrm{T}}(\tau)\right\rangle =\beta_{0}e^{\rmi\omega\tau}\left\vert \varepsilon_{0}^{(1)}\right\rangle +\sum_{k=1}^{M-1}\beta_{k}\left\vert \varepsilon_{k}^{(1)}\right\rangle ,
\end{equation}
where $\beta_{k}$s are the expanding coefficients in the basis $\left\{ \left\vert \varepsilon_{k}^{(1)}\right\rangle \right\} $.

Following the end of the pulse, the laser field is switched off and
the system then evolves free of laser pulse influence, yet still follows
the CMRT master equation describing population relaxation and dephasing
in the exciton basis. In this case, since $\left\vert \varepsilon_{k}^{(2)}\right\rangle $
are not the eigen states of $H_{e}^{(1)}$, we must expand the space
of jump states and initialize $M$ new states, i.e., for $k=0,1,\cdots M-1$,
\begin{eqnarray}
\left\vert \psi_{k+M+1}(\tau)\right\rangle  & = & U(\tau)\left\vert \varepsilon_{k}^{(2)}\right\rangle =C_{k0}^{(2)}e^{\rmi\omega\tau}\left\vert \varepsilon_{0}^{(1)}\right\rangle +\sum_{k^{\prime}=1}^{M-1}C_{kk^{\prime}}^{(2)}\left\vert \varepsilon_{k^{\prime}}^{(1)}\right\rangle ,
\end{eqnarray}
Thus, to describe both laser-induced dynamics and dissipative dynamics
of an excitonic system with $M$ chromophores, we must consider $2M+1$
possible states for the system to propagate in, i.e., $\left\vert \varepsilon_{k}^{(1)}\right\rangle $
for $k=0,1,\cdots M-1$, and $\left\vert \psi_{j}(t)\right\rangle $
for $j=M,M+1,\cdots2M$. The corresponding numbers of ensemble members
in the states $N_{\alpha}^{(\mathrm{II})}$ ($\alpha=0,1,\ldots2M$)
are determined by the NMQJ method to solve the CMRT master equation
(\ref{eq:OriEq}) with the initial values given as\begin{equation}N_{\alpha}^{(\mathrm{II})}(\tau)= \cases{0,&\textrm{for} $\alpha=0,1,\cdots M-1$,\\N_{M}^{(\mathrm{I})}(\tau),& \textrm{for} $\alpha=M$,\\N_{\alpha-M-1}^{(\mathrm{I})}(\tau),&\textrm{for} $\alpha=M+1,M+2,\cdots 2M$.}\end{equation}Finally,
the density matrix can be calculated from the time-dependent distribution
of numbers of ensemble members in states $N_{\alpha}^{(\mathrm{II})}(t)$
as 
\begin{equation}
\rho(t)=\frac{1}{N}\left[\sum_{k=0}^{M-1}N_{k}^{(\mathrm{II})}(t)\left\vert \varepsilon_{k}^{(1)}\right\rangle \left\langle \varepsilon_{k}^{(1)}\right\vert +\sum_{j=M}^{2M}N_{j}^{(\mathrm{II})}(t)\left\vert \psi_{j}(t)\right\rangle \left\langle \psi_{j}(t)\right\vert \right].
\end{equation}

We remark that the above procedure can be generalized to simulate
arbitrary pulse shape as long as the laser pulses only induce transitions
between the ground state and single-excitation states. In this case,
the eigen states in the rotating frame are always the superpositions
of the ground state and single-excitation states. Therefore, once
the instantaneous eigen states are changed due to the change in system-field
interaction, $M$ new trajectories will be initialized as the previous
eigen states are replaced by the eigen states of the new Hamiltonian.
In \ref{sec:appGAPS} we present the numerical details for simulating
the interaction between an excitonic system and a laser pulse with
arbitrary pulse profile.

The generalization of the combined CMRT-NMQJ method for a system in
the presence of laser fields can be easily extended to describe multiple
pulses to enable the simulation of $N$-wave-mixing optical experiments.
For example, the method described in this work can be applied to calculate
third-order response functions \cite{Mukamel:1999us}. Additionally,
since the combined CMRT-NMQJ approach describes the full laser-driven
dynamics of a dissipative excitonic system, it can be used together
with the density matrix based approach for the calculation of photon-echo
signals \cite{Cheng:2007dg,Gelin:2009fx} to evaluate nonlinear spectra,
including 2D electronic spectra and photon-echo peakshift signals.
Although 2D electronic spectra can not be calculated without 2-exciton
manifold, it would be straightforward to generalize the results presented
in this work to include dynamics in the two-exciton manifold \cite{Zhang:1998p3588,Cheng:2007dg}.
As a result, nonlinear spectra for coupled systems, which are often
difficult to calculate due to the need to average over static disorder,
can be calculated by using the combined CMRT-NMQJ approach. We emphasize
that in addition to the favorable system-size scaling due to dynamical
propagation in the wavefunction space, this approach benefits from
the fact that the average over static disorder can be performed simultaneously
with the average over quantum trajectories. Therefore, the combined
CMRT-NMQJ method should provide a much more numerically efficient
means for the evaluation of nonlinear spectra in large condensed-phase
systems. 

Note that in addition to the NMQJ method, quantum trajectory methods
such as the non-Markovian quantum state diffusion approach \cite{Diosi:1998um,Strunz:1999tj}
offer alternatives to the unraveling of quantum dynamics in light-harvesting
systems \cite{Rebentrost:2009kv,deVega:2011ih}. In that approach,
an integration of memory kernel and a functional derivative of the
wavefunction with respect to the noise are required to obtain the
stochastic Schr\"{o}dinger equation. Recently, de Vega has applied
the quantum state diffusion approach within the so-called post-Markov
approximation to describe non-Markovian dissipative dynamics of a
network mimicking a photosynthetic system \cite{deVega:2011ih}. However,
generally speaking it is not straightforward to unravel a non-Markovian
quantum master equation using the quantum state diffusion approach.
In this work we aim to focus on the CMRT method, which is a time-local
quantum master equation with a generalized Lindblad form. As a result
the quantum jump approach proposed by Piilo and coworkers seems to
be a natural way to implement the propagation of the time-local CMRT
equation. A detailed comparison of the numerical efficiency and domains
of applicability of the various quantum jump and quantum trajectory
based methods is given in Refs. \cite{Brun:2002p93736} and \cite{Piilo:2009p89261}.
We believe these methods could provide superior numerical performance
as well as useful physical insights for EET dynamics in complex and
large systems.

\section{EET Dynamics of FMO}

\label{sec:FMO}

In the previous sections, by combing the CMRT with the NMQJ method,
we have proposed an efficient approach to simulate the EET dynamics
of photosynthetic systems in the presence of laser fields. In order
to implement the NMQJ method, the master equation obtained by the
CMRT is revised in the Lindblad form. Since several approximations
are made in order to efficiently simulate the quantum dynamics of
a large quantum open system for a broad range of parameters, it is
necessary to verify the validity of this new approach for photosynthetic
systems. 

\begin{figure}
\centering\includegraphics[height=6cm]{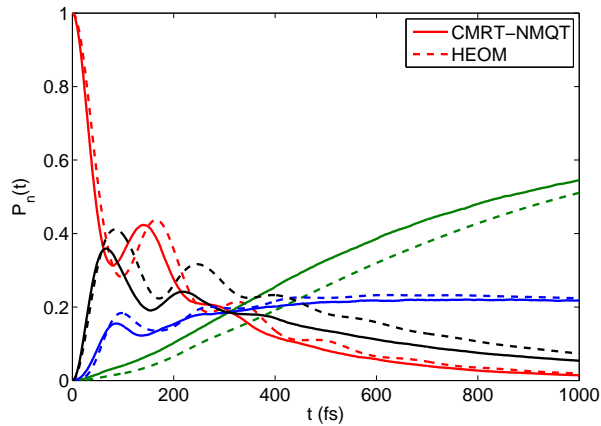}\caption{Site population dynamics of FMO with initial excitation on site $6$.
The simulation parameters are $T=77$K, $\lambda=35$cm$^{-1}$, and
the effective Hamiltonian used by Ishizaki and Fleming (cf. \cite{Ishizaki:2009p75287})
is used in both calculations. Solid curves are obtained by the CMRT-NMQJ
approach, and dashed curves depict results from the HEOM method. Red
curves are for site $6$, green curves for site $3$, blue curves
for site $4$, and black curves for site $5$. In order to make the
figure clear, population dynamics for other sites whose populations
always stay less than $0.1$ are not shown here.\label{fig:FMO}}
\end{figure}

As a numerical demonstration, in figure \ref{fig:FMO}, we compare
the population dynamics of EET in FMO simulated by two different approaches,
i.e., the combined CMRT-NMQJ approach and the numerically exact HEOM
method from Ishizaki and Fleming \cite{Ishizaki:2009p75287}. Clearly,
the results from the two methods are in good agreement. The combined
CMRT-NMQJ method does not only reproduce the coherent oscillations
in the short-time regime but is also consistent with the HEOM in the
long-time limit. In the short-time regime the oscillations are somewhat
suppressed in the case of CMRT-NMQJ approach. This is likely due to
the simplification of the pure-dephasing process by the direct projection
operations. If more realistic dephasing operators as implemented in
\cite{Makarov:1999cy} is applied to simulate the pure-dephasing process,
we would expect a better accuracy from the CMRT-NMQJ approach. However,
this additional effort may require much more computational time with
only marginal improvements for EET dynamics in photosynthetic systems.
Therefore, it is a reasonable tradeoff for numerical efficiency that
we adopt the Lindblad-form CMRT and make use of the NMQJ method to
simulate quantum dynamics of realistic photosynthetic systems

Note that because the NMQJ is a numerical propagation scheme, the
accuracy of the relaxation dynamics calculated by this method actually
depends on the Lindblad-form CMRT, which is effectively a generalization
to the widely used modified-Redfield theory in terms of population
dynamics. Modified-Redfield approach has been proven to provide excellent
results for biased-energy systems in photosynthesis \cite{Novoderezhkin:2010p82122},
therefore we expect the combined CMRT-NMQJ method described in this
work to be broadly applicable to EET in photosynthetic systems.

\section{Dimer Model}

To demonstrate the combined CMRT-NMQJ simulation of laser-driven dynamics,
we investigate the behavior of a model dimer system under the influence
of laser excitation in this section. Note that in previous simulations
of EET dynamics in photosynthetic systems the photoexcitation process
was often overlooked and the instantaneous preparation of a given
initial condition at zero time is normally assumed. Given that the
selective preparation of a specific excitonic state is a highly non-trivial
task (in particular the site-localized states often assumed in coherent
EET simulations), it is crucial that an experimentally realistic photoexcitation
process can be considered in a complete dynamical simulation \cite{Mancal:2010kc,Brumer:2012ib}.
Here we show that the combined CMRT-NMQJ approach is capable of describing
the excitation process. We further investigate the behavior of a model
dimer system (figure \ref{fig:Energy-Diagram-Pulsed}) under the influence
of laser excitation to study the interplay between the laser excitation
and ultrafast EET dynamics.

\subsection{Absorption Spectrum}

\begin{figure}
\centering\includegraphics[height=6cm]{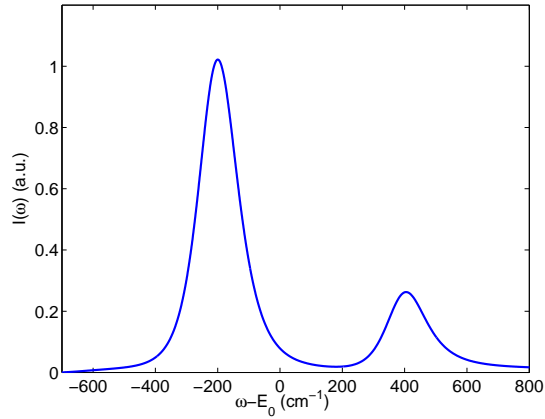}

\caption{The normalized absorption lineshape for the electronic coupling $J=300$cm$^{-1}$,
electronic transition dipole moments $\mu_{10}=10$D and $\mu_{20}=5$D,
reorganization energy $\lambda=35$cm$^{-1}$ and temperature $T=300$K.\label{fig:AbsLineShape}}
\end{figure}

Before we examine the laser-induced dynamics we demonstrate that the
absorption spectrum can be obtained by the CMRT method (cf. (\ref{eq:chi})).
In figure \ref{fig:AbsLineShape}, we show numerically calculated
absorption lineshape for dimer system with ground state energy $E_{0}=-12800$cm$^{-1}$,
site energies $E_{1}=120$cm$^{-1}$ and $E_{2}=100$cm$^{-1}$, electronic
coupling $J=300$cm$^{-1}$, and the electronic transition dipole
moments in the eigen basis are set to be $\mu_{10}=10$D and $\mu_{20}=5$D,
respectively. We further assume identical Ohmic baths described by
the spectral density 
\begin{equation}
J(\omega)=\lambda\frac{\omega}{\omega_{c}}e^{-\omega/\omega_{c}},\label{eq:Ohmic}
\end{equation}
with the reorganization energy $\lambda=35$cm$^{-1}$ and the cut-off
frequency $\omega_{c}=50$cm$^{-1}$, and temperature $T=300$K. The
simulated spectrum exhibits two peaks located at $\varepsilon_{2}^{(1)}-\varepsilon_{0}^{(1)}$
and $\varepsilon_{1}^{(1)}-\varepsilon_{0}^{(1)}$, corresponding
to the transitions from the ground state $\vert\varepsilon_{0}^{(1)}\rangle$
to two exciton states $\vert\varepsilon_{1}^{(1)}\rangle$ and $\vert\varepsilon_{2}^{(1)}\rangle$. 

Figure \ref{fig:AbsLineShape} shows that the absorption lineshape
is well captured by the CMRT approach, which is important if a comparison
to experimental data is to be made \cite{Novoderezhkin:2006p516,Renger:el}.
The relative peak heights are given by the ratio of $\left|\mu_{k0}\right|^{2}$.
Here, the widths of the two peaks are nearly the same as the effect
of relaxation-induced broadening is not significant in this particular
parameter set. Physically speaking, the pure-dephasing process is
much faster than the energy relaxation from the high-energy state
$\left\vert \varepsilon_{2}^{(1)}\right\rangle $ to the low-energy
state $\left\vert \varepsilon_{1}^{(1)}\right\rangle $ when the dimer
system is excited. For other cases, the contribution from the relaxation-induced
broadening might play a more important role \cite{Novoderezhkin:2006p516}.

\subsection{Laser-induced Dynamics }

\begin{figure}
\centering\includegraphics[bb=0bp 0bp 440bp 358bp,clip,height=6cm]{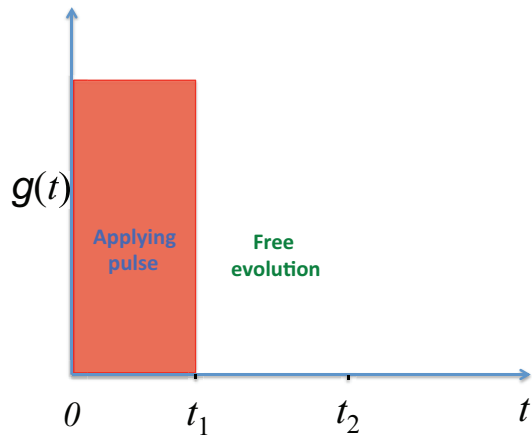}\caption{Time sequence of our scheme. The laser is turned on during the period
$\left[0,t_{1}\right]$. Afterwards, the laser is turned off and the
system is left to evolve free of control.\label{fig:Time-Sequence}}
\end{figure}

In this section, we apply the CMRT-NMQJ approach to investigate the
quantum dynamics of a dimer system after laser excitation. As shown
in figure \ref{fig:Time-Sequence}, the dimer system initially in
the ground state is applied with a square laser pulse during the time
period $\left[0,t_{1}\right]$. Note that a square pulse is used here
for the sake of simplicity (see \ref{sec:appGAPS}). Due to the coupling
induced by the laser pulse, the dimer is coherently excited into a
superposition between the ground state and the single-excitation states.
After the laser is turned off at $t=t_{1}$, the dimer is left alone
to evolve under the influence of system-bath couplings. The energy
diagrams for the case with and without a laser pulse are depicted
in figures \ref{fig:Energy-Diagram-Pulsed} and \ref{fig:Energy-Diagram-Free},
respectively.

In our numerical simulation, we adopt the following parameters to
model photosynthetic systems: the ground state energy $E_{0}=-12800$cm$^{-1}$,
the site energies $E_{1}=200$cm$^{-1}$ and $E_{2}=100$cm$^{-1}$.
Moreover, we consider two specific cases $J=120$cm$^{-1}$ and $J=20$cm$^{-1}$
to explore the electronic coupling's influence on the dynamics. We
further assume identical Ohmic spectral densities (cf. (\ref{eq:Ohmic}))
for each site with a reorganization energy of $\lambda=35$cm$^{-1}$,
cut-off frequency $\omega_{c}=50$cm$^{-1}$, and temperature $T=300$K.
In all cases, the laser duration $t_{1}$ is set to $100$fs. In addition,
in order to make the results converge within a reasonable computational
time, we use a moderate time step $dt=1$fs and average over a sufficiently-large
number of trajectories, i.e., $N=10^{5}$.

\subsubsection{Laser-induced Population dynamics}

\begin{figure}
\includegraphics[height=6cm]{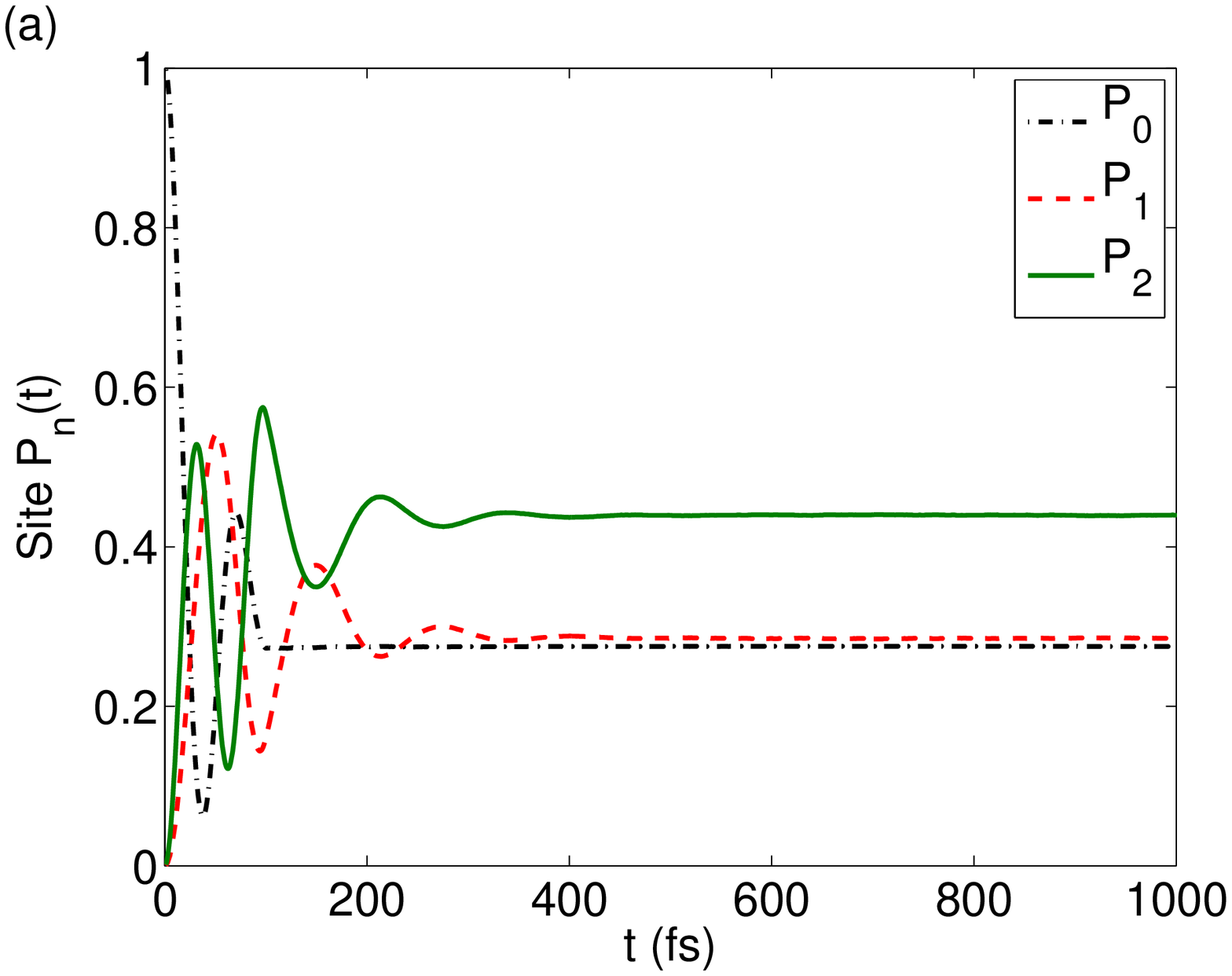}\includegraphics[height=6cm]{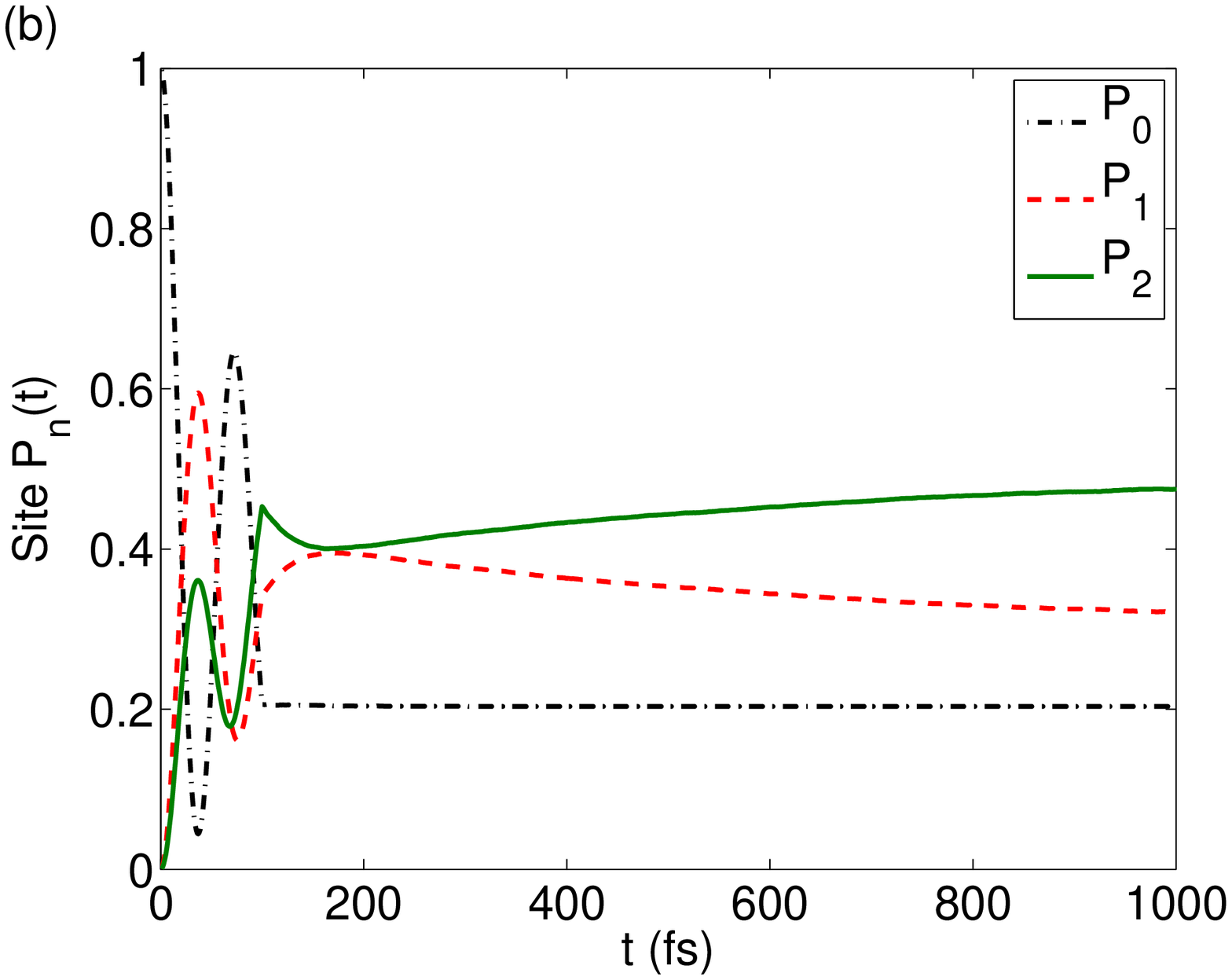}\caption{Time evolution of site populations $P_{n}(t)$ for dimer systems interacting
with a near-resonant field (a) intradimer electronic coupling $J=120$cm$^{-1}$,
(b) $J=20$cm$^{-1}$. We set the field-induced couplings between
the ground state and the two eigen states, $g_{10}=100$cm$^{-1}$
and $g_{20}=200$cm$^{-1}$. \label{fig:Pn-Near-Resonant}}
\end{figure}

Figures \ref{fig:Pn-Near-Resonant}(a) and \ref{fig:Pn-Near-Resonant}(b)
show the time-evolution of the site populations initialized by a laser
pulse for strongly ($J=120$cm$^{-1}$) and weakly ($J=20$cm$^{-1}$)
electronically coupled dimers, respectively. The laser carrier frequency
is tuned at $\omega=13000$cm$^{-1}$, close to both transition energies.
Therefore we expect the laser pulse to induce significant coherence
between the two exciton states. As a result, during the pulse duration
(from $t=0$fs to $t=100$fs), the populations on both sites oscillate
with large amplitudes, for the laser field is nearly on resonance
with the transitions between the ground state $\vert\varepsilon_{0}^{(1)}\rangle$
and two eigen states $\vert\varepsilon_{1}^{(1)}\rangle$ and $\vert\varepsilon_{2}^{(1)}\rangle$.
After the laser is turned off at $t=100$fs, the population on the
ground state remains invariant since there is no relaxation from the
single-excitation states to the ground state in our model (cf. (\ref{eq:HSB})),
however, dynamics in the single-exciton manifold show strong $J$
dependence. A comparison between figures \ref{fig:Pn-Near-Resonant}(a)
and \ref{fig:Pn-Near-Resonant}(b) shows that although both systems
evolve to reach a thermal equilibrium due to the system-bath couplings,
the relaxation dynamics are qualitatively different due to the different
electronic coupling strengths. In the case with strong electronic
coupling (figure \ref{fig:Pn-Near-Resonant}(a)) coherent oscillations
persist for up to $400$fs, and the system then quickly relaxes to
the equilibrium. Clearly, the coherent energy relaxation is extremely
efficient in this case and the photoexcitation dynamics and the coherent
dynamics are intertwined \cite{Chin:2010p81675,Ai:2013wl}. We conclude
that in strongly coupled systems the nature of the photoexcitation
process must be explicitly considered in order to reasonably describe
the photon-induced dynamics, as a fictitiously assumed initial condition
prepared instantaneously at $t=0$ could not correctly capture the
photoexcitation process. In contrast, for the weak electronic coupling
case in figure \ref{fig:Pn-Near-Resonant}(b), the coherent oscillations
stop immediately after the pulse is turned off, and the system relaxes
incoherently and also slowly. In this case a clear time-scale separation
exists and a specific consideration of the photoexcitation process
may not be necessary. 

\begin{figure}
\includegraphics[height=6cm]{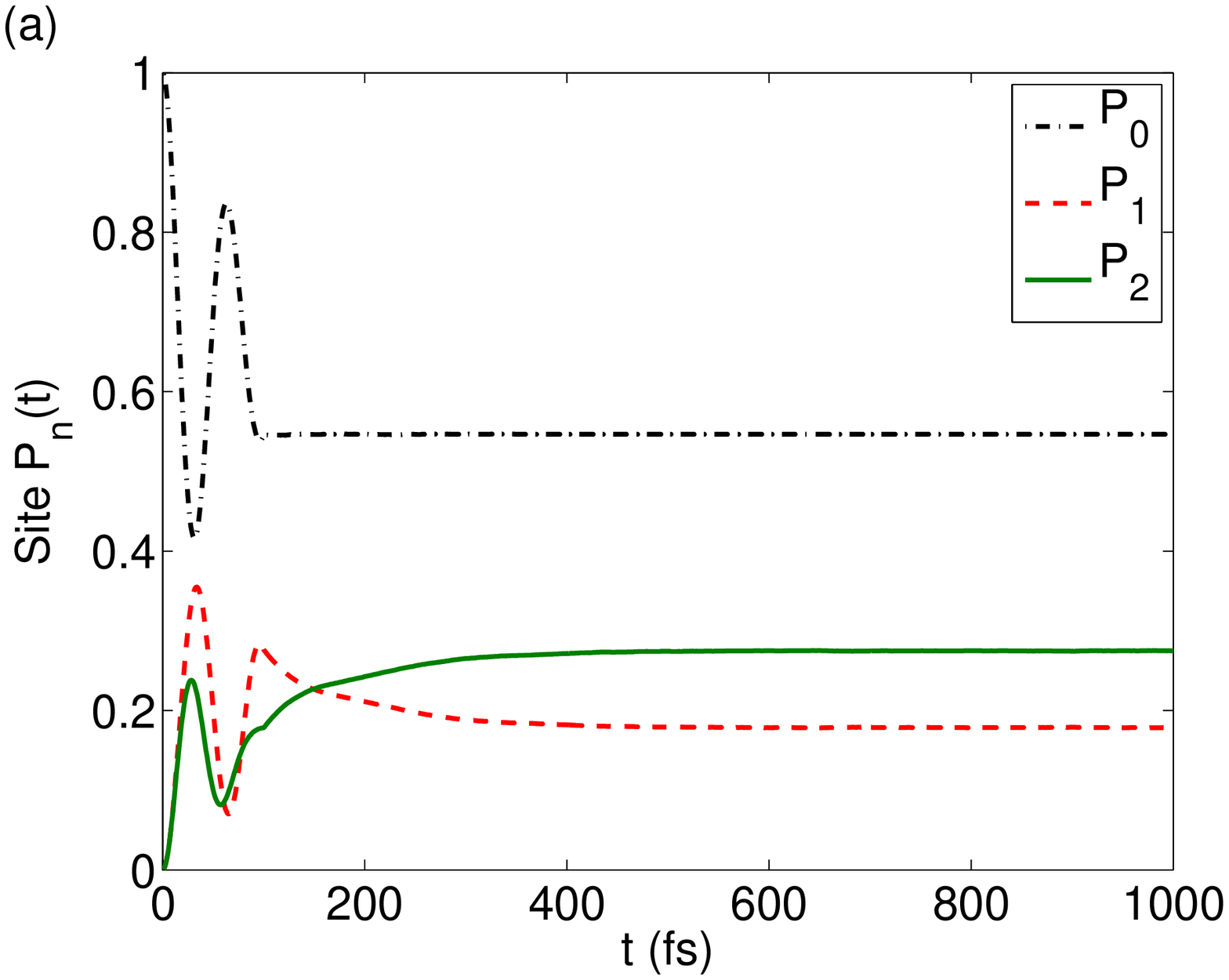}\includegraphics[height=6cm]{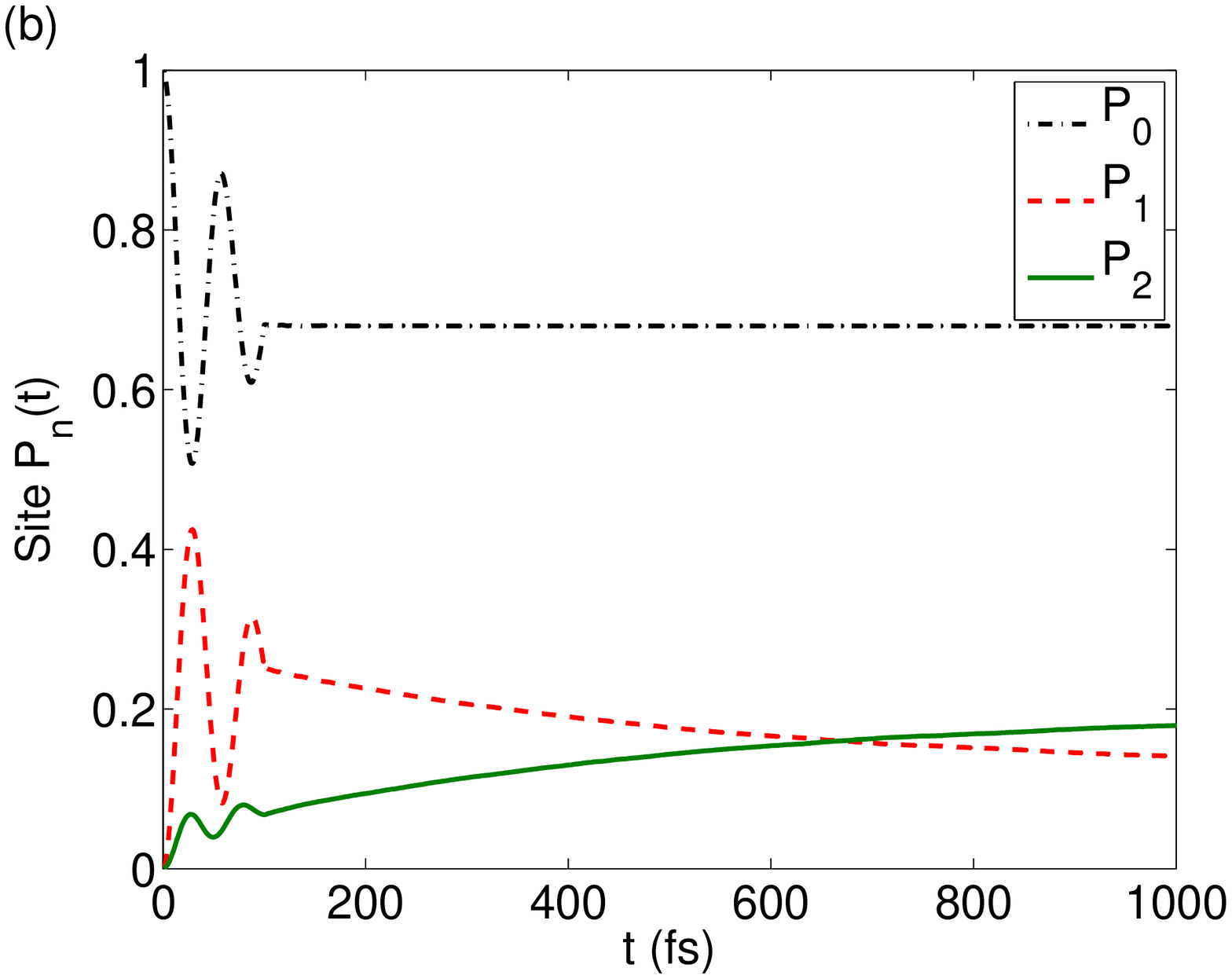}\caption{Time evolution of site populations $P_{n}(t)$ for dimer systems interacting
with a detuned field with electronic coupling (a) intradimer electronic
coupling $J=120$cm$^{-1}$, (b) $J=20$cm$^{-1}$. We set the field-induced
couplings between the ground state and the two eigen states, $g_{10}=40$cm$^{-1}$
and $g_{20}=200$cm$^{-1}$. The laser frequency is set at $\omega=13400$cm$^{-1}$
to be in near resonance with the transition between $\vert\varepsilon_{0}^{(1)}\rangle$
and $\vert\varepsilon_{2}^{(1)}\rangle$ but largely-detuned to the
transition between $\vert\varepsilon_{0}^{(1)}\rangle$ and $\vert\varepsilon_{1}^{(1)}\rangle$.\label{fig:Pn-Large-Detuning}}
\end{figure}

To investigate the effects of laser excitation condition, we explore
the effect of laser detuning on the quantum dynamics in figure \ref{fig:Pn-Large-Detuning}.
In this case, the laser carrier frequency is tuned at $\omega=13400$cm$^{-1}$.
Because the laser is largely detuned from the transition between $\vert\varepsilon_{0}^{(1)}\rangle$
and $\vert\varepsilon_{1}^{(1)}\rangle$ but nearly on resonance with
the transition between $\vert\varepsilon_{0}^{(1)}\rangle$ and $\vert\varepsilon_{2}^{(1)}\rangle$,
i.e., $g_{01}\ll\left|\varepsilon_{1}^{\prime}-\left(\varepsilon_{0}^{\prime}+\omega\right)\right|$
and $g_{02}\sim\left|\varepsilon_{2}^{\prime}-\left(\varepsilon_{0}^{\prime}+\omega\right)\right|$,
we expect strong Rabi oscillations for the $\vert\varepsilon_{0}^{(1)}\rangle\rightleftarrows\vert\varepsilon_{2}^{(1)}\rangle$
transition, and $\vert\varepsilon_{1}^{(1)}\rangle$ can not be effectively
excited when the pulse is present \cite{Scully:1997aa}. Indeed, figure
\ref{fig:Pn-Large-Detuning}(b) shows that in this weak electronic
coupling case, site $1$ is selectively excited within the pulse duration,
since it is the higher-energy site in our dimer model. Consequently,
figure \ref{fig:Pn-Large-Detuning}(b) indicates that specific initial
state, e.g., a specific site being excited, may be prepared if a proper
pulse is applied for an appropriate duration. In contrast, figure
\ref{fig:Pn-Large-Detuning}(a) shows that the oscillation amplitudes
of site populations, $P_{1}$ and $P_{2}$, are nearly the same in
the strong electronic coupling case, as in this case both sites contribute
significantly to the delocalized exciton state $\vert\varepsilon_{2}^{(1)}\rangle$. 

Noticeably, the excitation-relaxation dynamics presented in figures
\ref{fig:Pn-Near-Resonant}(a) and \ref{fig:Pn-Large-Detuning}(a)
are markedly different, i.e. no coherent evolution in figure \ref{fig:Pn-Large-Detuning}(a)
after $t=100$fs. Clearly, the dynamical behavior of the strongly
coupled dimer depends critically on the excitation conditions. Our
results further emphasize the notion that the photoexcitation process
must be explicitly considered \cite{Mancal:2010kc,Brumer:2012ib}.
Note that we aim to demonstrate the validity of the combined CMRT-NMQJ
method developed in this work. It is clear that much more efforts
must be paid to elucidate the effects of light source in natural light-harvesting
processes. In this regard, the methodology developed in this work
appears to be valuable for the study of coherent EET dynamics and
quantum coherent control in photosynthetic systems.

\subsubsection{Entanglement dynamics}

In the population dynamics, coherent evolutions are present when the
laser excitation or the electronic coupling is strong. To more effectively
follow coherent EET dynamics, it has been shown that entanglement
is a excellent probe for coherent dynamics in photosynthetic systems
\cite{Caruso:2009kb,Caruso:2010kb,Fassioli:2010kb,HosseinNejad:2010em,Ishizaki:2010ft,Liao:2010fh,Sarovar:2010hs}.
Here, we investigate the time-evolution of entanglement in the dimer
system as a demonstration of the combined CMRT-NMQJ method for coherent
EET dynamics. Because each pigment in the dimer is a natural qubit,
we especially focus on the evolution of concurrence \cite{Wootters:1998wt}.
For a two-qubits system with arbitrary density matrix $\rho$, the
concurrence is defined as \cite{Wootters:1998wt}
\begin{equation}
C\equiv\textrm{max}\left(0,\lambda_{1}-\lambda_{2}-\lambda_{3}-\lambda_{4}\right),
\end{equation}
where $\lambda_{j}$s are the square roots of eigen values of the
matrix $\rho\tilde{\rho}$ in the descending order. Here, 
\begin{equation}
\tilde{\rho}=\left(\sigma_{y}\otimes\sigma_{y}\right)\rho^{*}\left(\sigma_{y}\otimes\sigma_{y}\right)
\end{equation}
with $\rho^{*}$ being the complex conjugate of the density matrix
$\rho$. For a purely classical system, its concurrence vanishes,
while it would be unity for the maximum entangled states, e.g., Bell
states. For other states, the concurrence monotonously increases as
the state is more quantum-mechanically entangled. An analytical expression
that yields $C(t)$ from the matrix elements of $\rho(t)$ can be
derived. In our model, since we restrict the evolution of the system
within the sub-space without the double-excitation state, the density
matrix is of the block type
\begin{equation}
\rho=\left(\begin{array}{cccc}
0 & 0 & 0 & 0\\
0 & \rho_{22} & \rho_{21} & \rho_{20}\\
0 & \rho_{12} & \rho_{11} & \rho_{10}\\
0 & \rho_{02} & \rho_{01} & \rho_{00}
\end{array}\right),
\end{equation}
where the basis states are $\left|0\right\rangle =\left|g\right\rangle _{1}\left|g\right\rangle _{2}$,
$\left|1\right\rangle =\left|e\right\rangle _{1}\left|g\right\rangle _{2}$,
$\left|2\right\rangle =\left|g\right\rangle _{1}\left|e\right\rangle _{2}$,
and $\left|3\right\rangle =\left|e\right\rangle _{1}\left|e\right\rangle _{2}$.
Thus, the concurrence is simplified as
\begin{eqnarray}
C(\rho) & = & \max\left\{ 0,\sqrt{\rho_{11}\rho_{22}}+\left|\rho_{21}\right|-\left(\sqrt{\rho_{11}\rho_{22}}-\left|\rho_{21}\right|\right)\right\} \nonumber \\
 & = & 2\left|\rho_{21}\right|.
\end{eqnarray}
Because $\rho_{11}+\rho_{22}\geq2\left|\rho_{21}\right|$, the more
population is excited to the single-excitation states, the larger
the concurrence could be. In addition, at some time, when $\rho_{21}$
vanishes, we would expect the system evolves into the disentangled
state as the concurrence disappears.

\begin{figure}
\includegraphics[height=6cm]{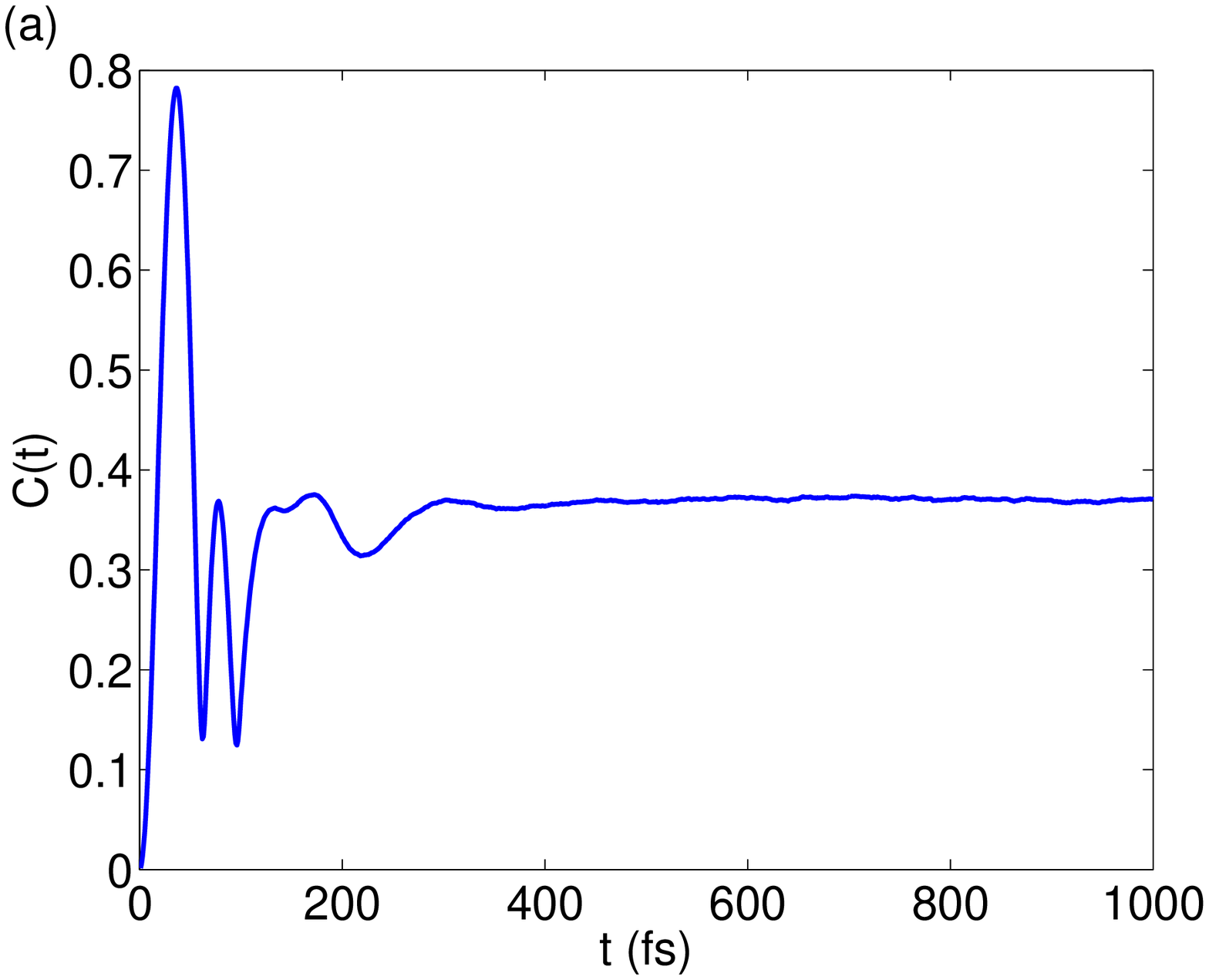}\includegraphics[height=6cm]{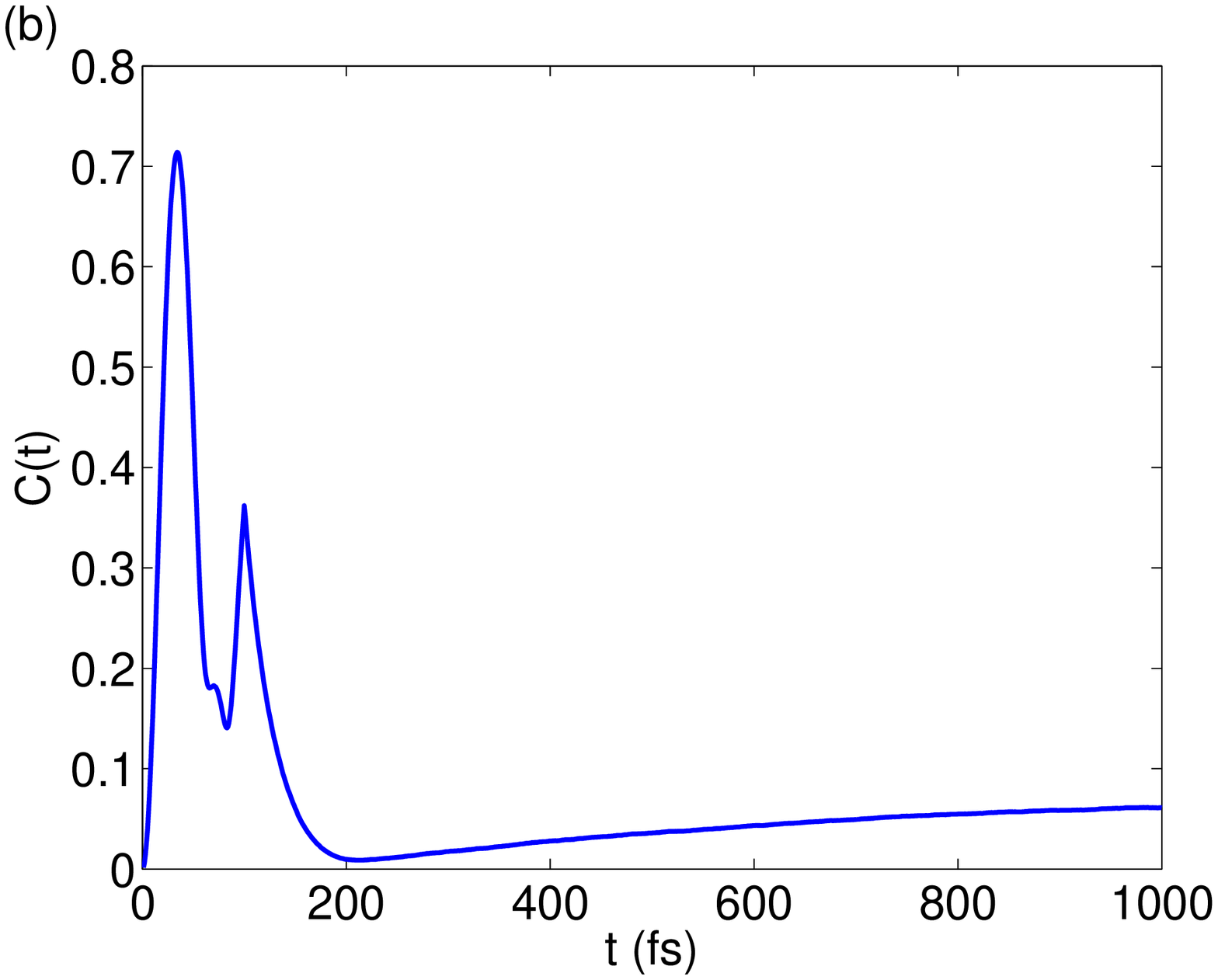}\caption{Time evolution of concurrence $C(t)$ for dimer systems interacting
with a near-resonant pulse. (a) intradimer electronic coupling $J=120$cm$^{-1}$,
(b) $J=20$cm$^{-1}$. Other parameters are the same as those in figure
\ref{fig:Pn-Near-Resonant}. \label{fig:Con-Near-Resonant}}
\end{figure}

In figure \ref{fig:Con-Near-Resonant}, we plot concurrence dynamics
for the condition that the excitation pulse is in near resonance with
both transitions. For both strong and weak electronic coupling cases,
there exists significant entanglement induced by the optical excitation.
However, the concurrence dynamics after the laser is switched off
at $100$fs are markedly different. Figure \ref{fig:Con-Near-Resonant}(a)
shows that for the strongly coupled dimer the concurrence quickly
evolves to reach a steady state with a considerable value as a result
of the large electronic coupling. In contrast, for the weak electronic
coupling case shown in figure \ref{fig:Con-Near-Resonant}(b), the
concurrence rapidly decays to near zero and then follows by a continuous
but slow rise. These results are consistent with the characteristics
of population dynamics as shown in figure \ref{fig:Pn-Near-Resonant},
indicating that the concurrence provides an effective tool to describe
coherent energy transfer process.

\begin{figure}
\includegraphics[height=6cm]{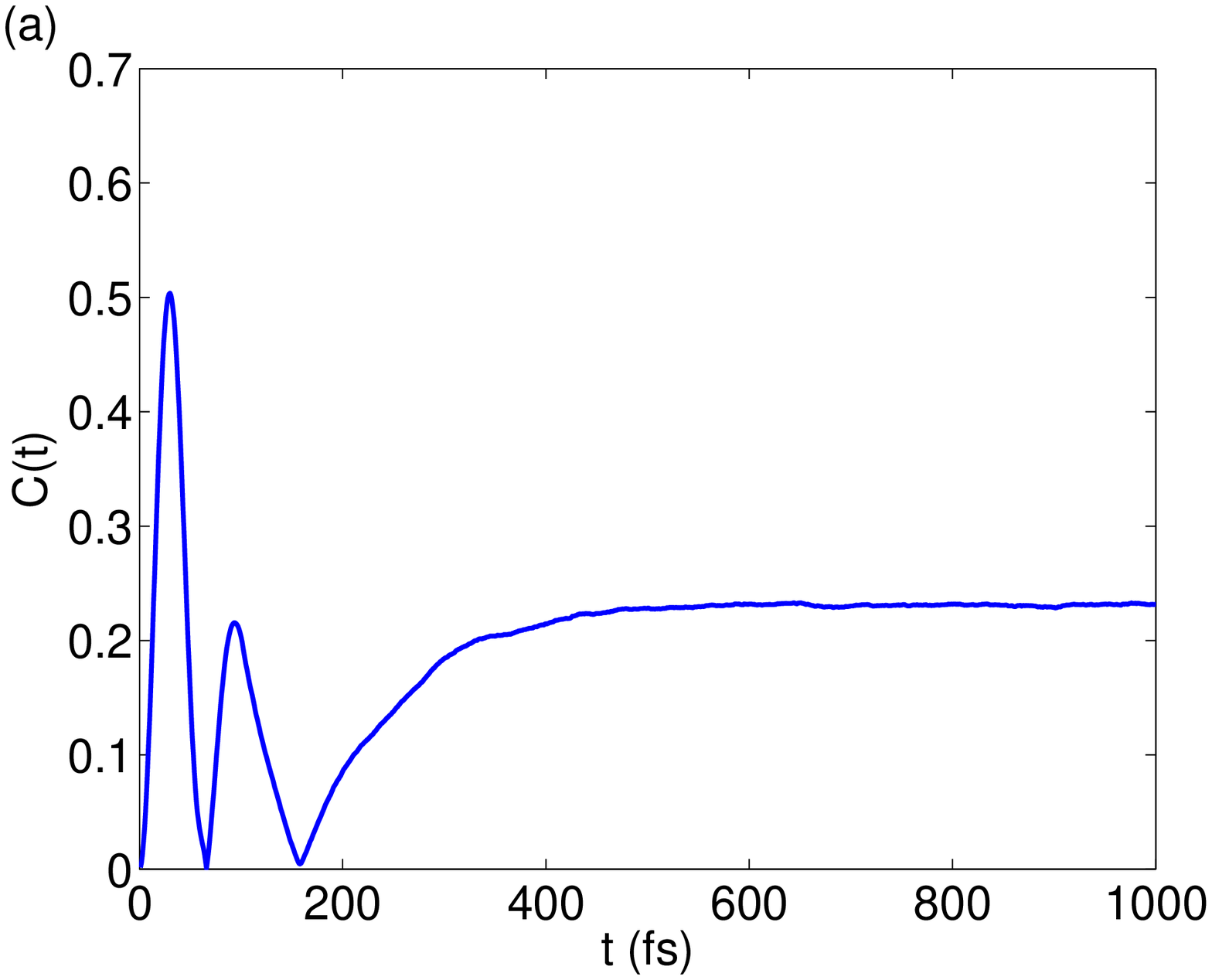}\includegraphics[clip,height=6cm]{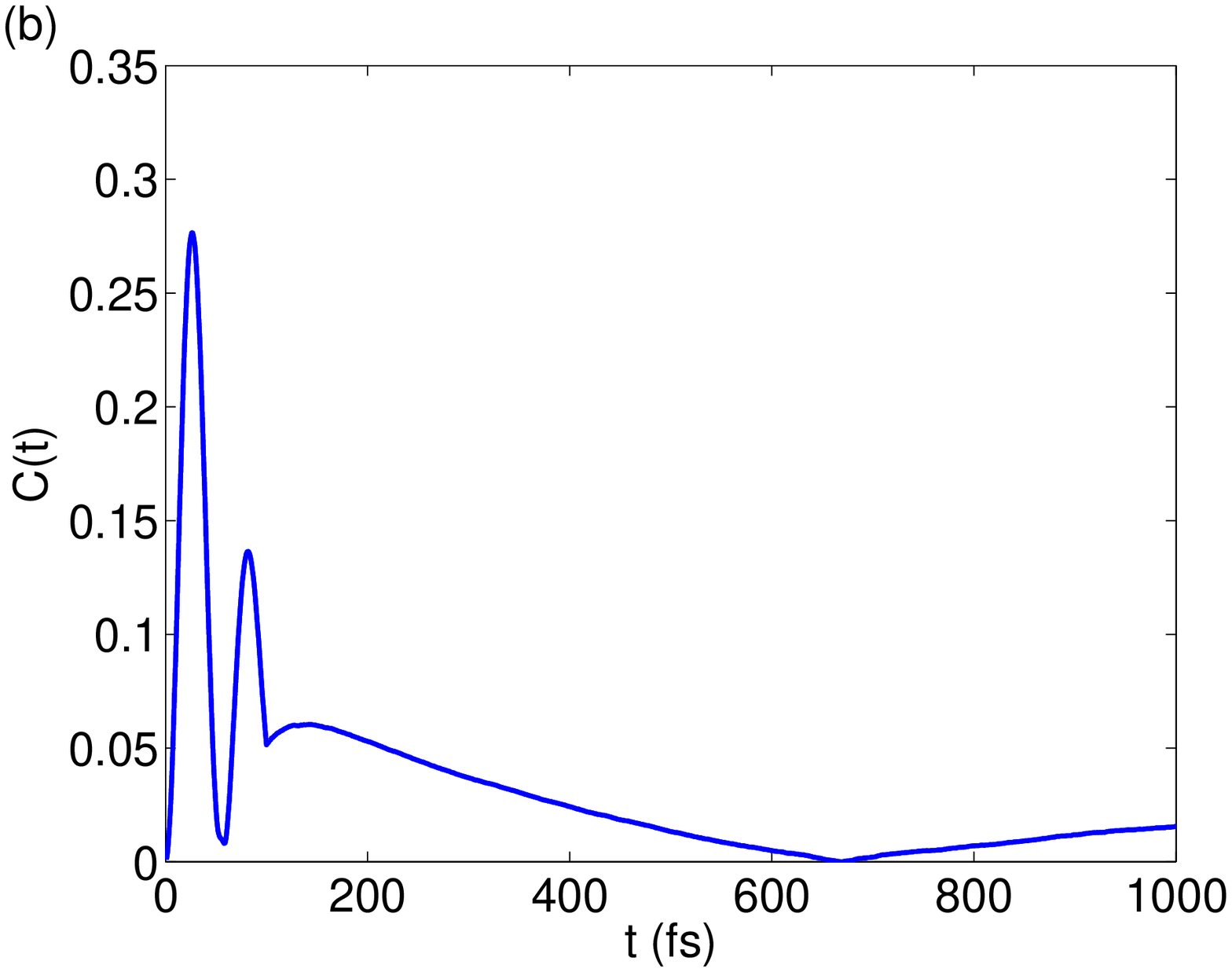}\caption{Time evolution of concurrence $C(t)$ for dimer systems interacting
with a detuned pulse. (a) intradimer electronic coupling $J=120$cm$^{-1}$,
(b) $J=20$cm$^{-1}$. Other parameters are the same as those in figure
\ref{fig:Pn-Large-Detuning}. \label{fig:Con-Large-Detuning}}
\end{figure}

Moreover, the evolution of concurrence with only one exciton state
selectively excited is shown in figure \ref{fig:Con-Large-Detuning}.
In the duration of pulse, the maximum entanglement generated by the
field is significantly smaller than that by a resonance field. This
is because less coherence between the two transitions can be created
as a consequence of the laser de-tuning. In addition, because the
laser is only in close resonance with one of the transitions, as shown
in figure \ref{fig:Con-Large-Detuning}(a), the concurrence at the
steady state is less than that achieved in figure \ref{fig:Con-Near-Resonant}(a),
as less population can be excited to the single-excitation states
by the weaker pulse. This discovery again emphasizes the importance
of the state preparation by photoexcitation.

Besides, we notice that for all cases there is an abrupt change at
$t=100$fs due to the discontinuity of the system-field interaction,
which is an artifact of the step-function pulse shape. We remark that
this problem can be solved by using a smooth pulse profile. In a realistic
experiment, a gaussian pulse should be applied to the system. The
combined CMRT-NMQJ approach can be generalized to consider a gaussian
pulse by decomposing it into a series of rectangular sub-pulses with
equal width (see details in \ref{sec:appGAPS}). Again, these results
indicate that the photoexcitation process plays an important role
in the induction of initial coherence in excitonic systems, therefore
system-field couplings must be explicitly considered in order to elucidate
electronic quantum coherence effects in photosynthesis \cite{Brumer:2012ib,Ishizaki:2012kf}.

\section{Conclusions}

In this paper, we have combined the CMRT \cite{YuHsienHwangFu:2012we}
approach and the NMQJ method \cite{Piilo:2008di,Piilo:2009p89261}
to develop a formalism that provides efficient simulations of coherent
EET dynamics of photosynthetic complexes under the influence of laser
fields. In order to implement the NMQJ propagation of the CMRT master
equation, the original CMRT master equation is revised in the Lindblad
form. In addition, the NMQJ approach is generalized to be suitable
for the case with a time-dependent Hamiltonian due to interactions
with laser fields. These new developments allow the efficient calculation
of quantum dynamics of photosynthetic complexes in the presence of
laser fields.

To demonstrate the effectiveness and efficiency of this new approach,
we apply the CMRT-NMQJ approach to simulate the coherent energy transfer
dynamics in the FMO complex, and compare the result to the one obtained
by the numerically exact HEOM method. We show that both results are
consistent in the long-time and short-time regimes. Furthermore, we
investigate photon-induced dynamics in a dimer system. For strongly
coupled dimers, coherent oscillations are observed in the population
dynamics during the pulse duration as well as the free evolution as
predicted by theory. In addition to the quantum dynamics in the excited
states, we also consider coherent effects induced by the laser detuning
in the stage of state preparation. We show that the dynamical behavior
of a strongly coupled dimer depends critically on the excitation conditions,
which further emphasizes the point that the photoexcitation process
must be explicitly considered in order to properly describe photon-induced
dynamics in photosynthetic systems. In addition, we investigate the
evolution of concurrence of a dimer in the presence of laser fields.
These results demonstrate the combined CMRT-NMQJ approach is capable
of simulation of EET in photosynthetic complexes under the influence
of laser control. Note that although the phases of the fields are
fixed in our calculations, the combined CMRT-NMQJ method is fully
capable of describing quantum coherent control phenomena by shaping
phases of the laser fields.

Compared to other numerical simulation methods, this new approach
has several key advantages. First of all, because it is based on the
CMRT, its validity in a broad range of parameters has been well characterized
\cite{Novoderezhkin:2010p82122,YuHsienHwangFu:2012we}, which is confirmed
in simulating the energy transfer dynamics of the FMO complex. Note
that the CMRT might fail in the regime where the exciton basis is
not a suitable choice \cite{Novoderezhkin:2010p82122}, e.g., $\lambda>J>\Delta\varepsilon$.
In that case, a combined Redfield-F\"{o}rster picture could be used
to provide accurate simulations of EET dynamics \cite{Novoderezhkin:2011ha,Bennett:2013gi}.
Because the generalized F\"{o}rster dynamics only add diagonal population
transfer terms between block-delocalized exciton states, which directly
fit into a Lindblad form, an extension of the methods describe here
to simulate EET dynamics in the Redfield-F\"{o}rster picture will
be straightforward. Second, as this approach incorporates the NMQJ
method, it effectively reduces the calculation time \cite{Piilo:2008di}
and therefore is capable of simulating the quantum dynamics in large
photosynthetic systems. For an $M$-level system in the absence of
laser fields, the effective ensemble size $N_{\mathrm{eff}}=M+1$
in the CMRT-NMQJ approach can be sufficiently smaller than that needed
in the non-Markovian quantum state diffusion, e.g. $N=10^{4}$ for
a two-level system in Ref. \cite{Strunz:1999tj}. In addition, the
problem of positivity violation, which may be encountered by master
equation approaches, can be resolved by turning off the quantum jump
once the population of the source state vanishes. Third, it can make
use of parallel computation, because the sampling of trajectories
can be efficiently parallelized, and it can take the average over
static disorder at the same time as the average over trajectories.
Finally, since the absorption spectrum \cite{Novoderezhkin:2006p516}
and other nonlinear optical signals should be efficiently simulated
within the same theoretical framework, all the parameters used in
the simulation can be self-consistently obtained by fitting the experimental
data using this theoretical approach.

Our motivation to extend the CMRT theory to include time-dependent
fields and NMQJ propagation is to provide a tool for the study of
quantum control of photoexcitation and energy flow. In order to simulate
EET dynamics in a coherent control experiment, it is necessary to
consider a time-dependent Hamiltonian that includes the influence
of control laser fields. Quantum coherent control lies at the heart
of human's exploration in the microscopic world \cite{Brif:2010jr,Shapiro:2003aa,Yan:2012dj}.
Its amazing power manifests in manipulating quantum dynamics by systematic
control design methods \cite{Dong:2010wk,DAlessandro:2007vh,Wiseman:2010vw,Mabuchi:2005ek}.
The strong motivation to apply quantum coherent control to EET lies
in not only the aspiration to knowledge but also the energy need to
learn from the highly-efficient photosynthetic systems to improve
artificial light harvesting. We believe the methodologies presented
in this work would be useful for further investigation of using coherent
quantum control methods to direct energy flow in photosynthetic as
well as artificial light-harvesting systems. For example, we are currently
using the theoretical framework described in this work to investigate
how laser pulse-shape and -phase affect EET dynamics in photosynthetic
systems.

\ack{}{We thank stimulating discussions with Jian Ma, Yao Shen and Yi-Hsien
Liu. YCC thanks Akihito Ishizaki for kindly providing the HEOM results
for FMO. QA thanks the National Science Council, Taiwan (Grant No.
NSC 101-2811-M-002-148) for financial support. YCC thanks the National
Science Council, Taiwan (Grant No. NSC 100-2113-M-002-008-MY3), National
Taiwan University (Grant No. 103R891305), and Center for Quantum Science
and Engineering (Subproject: 103R891401) for financial support. We
are grateful to Computer and Information Networking Center, National
Taiwan University and the National Center for High-performance Computing,
Taiwan for the support of high-performance computing facilities. }

\appendix

\section{NMQJ Propagation with Arbitrary Pulse Shapes\label{sec:appGAPS}}

In Sec. \ref{sec:CMRTCP}, we presented the revised CMRT for the case
with a step-function pulse. However, in a realistic case, one more
often encounters a gaussian pulse. Because the NMQJ approach propagates
the dynamics in discrete time intervals, it is trivial to generalize
the short-time square pulse propagation scheme given in Sec. \ref{sec:CMRTCP}
to treat the influence of a laser pulse with arbitrary pulse profiles. 

\begin{figure}
\centering\includegraphics[bb=0bp 0bp 544bp 430bp,clip,width=7cm]{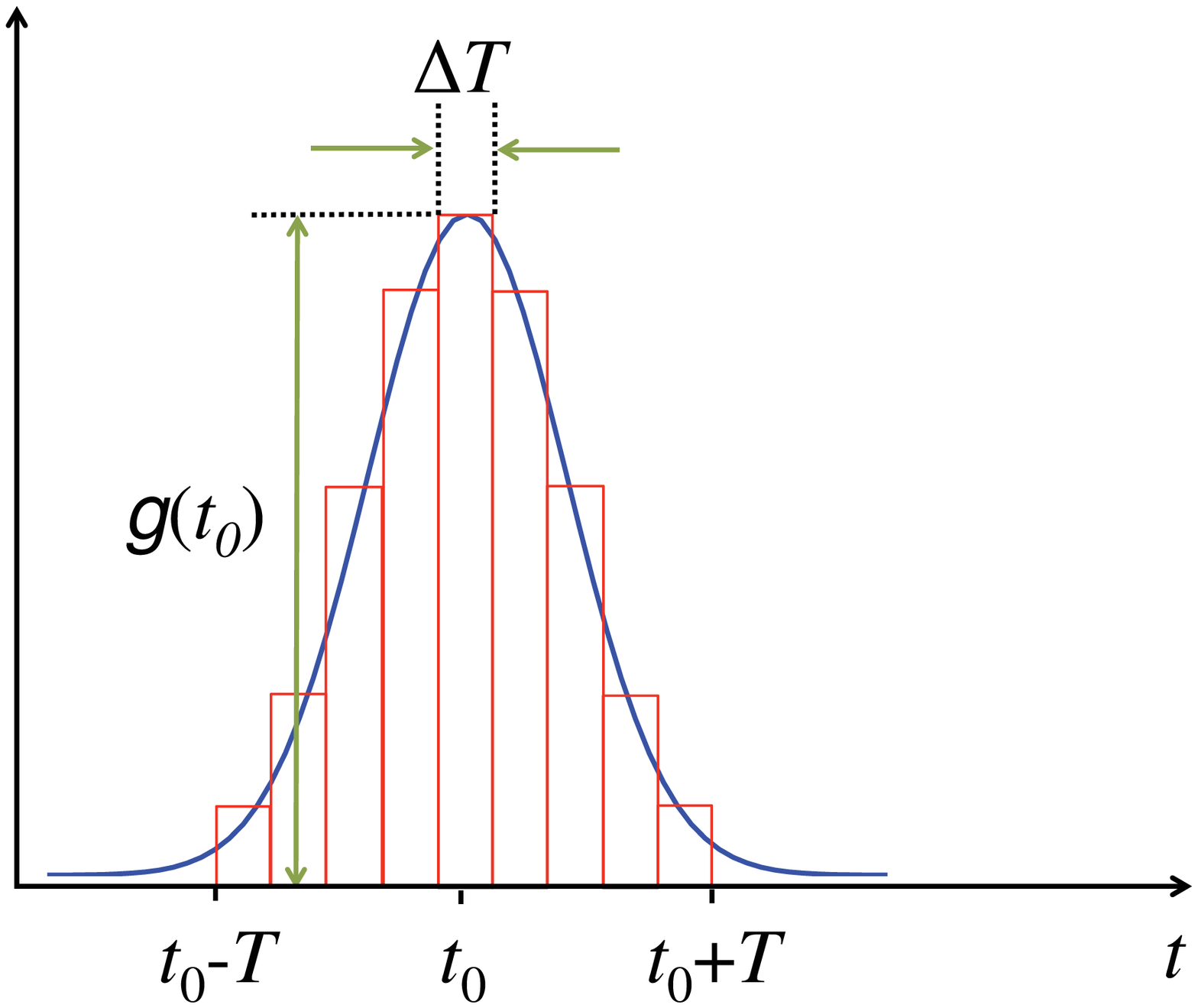}

\caption{Schematic diagram for decomposing a gaussian pulse centered at $t_{0}$
with width $T$ into $N_{p}$ rectangular sub-pulses with equal width
$\Delta T$.\label{fig:Gaussian}}
\end{figure}

In figure \ref{fig:Gaussian}, we schematically demonstrate how to
decompose a gaussian pulse with profile $g(t)=g_{0}\exp\left[-2\left(t-t_{0}\right)^{2}/T^{2}\right]$
into $N_{p}$ rectangular sub-pulses with equal width $\Delta T$.
The height of each sub-pulse is set to the strength of the gaussian
pulse right at the middle of the sub-pulse. The width $\Delta T$
is determined in such a way that by increasing the number of sub-pulses
and keeping the duration $[t_{0}-T,t_{0}+T]$ to be fixed, i.e., $N_{p}=2^{n}+1$
($n=1,2,\cdots$) and $N_{p}\Delta T=2T$, the area difference between
the gaussian pulse and the total summation of all sub-pulses does
not exceed a pre-determined accuracy threshold. During the time propagation,
when the Hamiltonian switches to the case with a different sub-pulse,
$M$ new trajectories are added to the propagation. At the end of
the sub-pulse sequence, there will be $N_{p}M$ states to propagate
in the simulation, which still maintains the linear scaling of the
method. For instance, when the gaussian pulse is decomposed into $N_{p}=513$
rectangular sub-pulses, the relative error between the areas of the
discretized pulse and the original Gaussian pulse is less than $10^{-6}$,
and the total number of states in the propagation only increases linearly
to $514M+1$. We remark that the above decomposition procedure holds
for arbitrary pulse shapes.

\section{Lindblad Form CMRT Mater Equation\label{sec:appELF}}

In this appendix, we show that (\ref{eq:OriEq}) can be recast into
the generalized Lindblad form. First of all, inspired by the discussion
in \cite{Piilo:2008di,Piilo:2009p89261}, we shall specifically show
that the pure-dephasing term in the CMRT can be rewritten in the Lindblad
form:
\begin{eqnarray}
-\sum_{k}\frac{\Gamma_{k}}{2}\left[\left\{ A_{kk}^{\dagger}A_{kk},\rho\right\} -2A_{kk}\rho A_{kk}^{\dagger}\right] & = & -\sum_{k\neq k^{\prime}}\frac{\Gamma_{k}+\Gamma_{k^{\prime}}}{2}\rho_{kk^{\prime}}\left\vert \varepsilon_{k}\right\rangle \left\langle \varepsilon_{k^{\prime}}\right\vert \nonumber \\
 & = & -\sum_{k\neq k^{\prime}}R_{kk^{\prime}}^{\textrm{pd}}\rho_{kk^{\prime}}\left\vert \varepsilon_{k}\right\rangle \left\langle \varepsilon_{k^{\prime}}\right\vert ,
\end{eqnarray}
where
\begin{equation}
R_{kk^{\prime}}^{\textrm{pd}}(t)=\frac{1}{2}\left(\Gamma_{k}+\Gamma_{k^{\prime}}\right).
\end{equation}
Here, we have made use of the properties of the diagonal jump operators,
i.e., 
\begin{equation}
A_{kk}^{\dagger}A_{kk}=A_{kk}^{\dagger}=A_{kk}.
\end{equation}
Clearly, the elements of the pure dephasing rate matrix, $R_{kk^{\prime}}^{\textrm{pd}}(t)$,
are real and symmetric with respect to the matrix diagonal. In addition,
the diagonal terms vanish, i.e.,\begin{equation}R_{kk^{\prime}}^{\textrm{pd}}= \cases{\textrm{real},&\textrm{for} $k\neq k^{\prime}$,\\0,&\textrm{for} $k=k^{\prime}$.}\end{equation}

Since there are $M(M-1)/$2 independent pure-dephasing rates $R_{kk^{\prime}}^{\textrm{pd}}$
but $M$ Lindblad-form dephasing rates $\Gamma_{k}$, we need some
additional constraints in order to fit $M$ Lindblad-form dephasing
rates $\Gamma_{k}$ from $M(M-1)/$2 independent pure-dephasing rates
$R_{kk^{\prime}}^{\textrm{pd}}$. In this work, we propose to use
a least-square fit to obtain the Lindblad-form dephasing rates. Therefore,
we require the mean-square displacement to be minimal with respect
to all the Lindblad rates:: 
\begin{equation}
\frac{\partial}{\partial\Gamma_{a}}\sum_{k=1}^{M-1}\sum_{k^{\prime}=k+1}^{M}\left[R_{kk^{\prime}}^{\textrm{pd}}-\frac{1}{2}\left(\Gamma_{k}+\Gamma_{k^{\prime}}\right)\right]^{2}=0,
\end{equation}
where $a=1,2,\ldots M$. After simplification, we obtain a system
of linear equations for the Lindblad-form dephasing rates, that is
\begin{eqnarray}
\frac{1}{2}\sum_{k=1}^{a-1}\Gamma_{k}+\frac{1}{2}(2M-a)\Gamma_{a}+\sum_{k=a+1}^{M-1}\Gamma_{k}+\frac{1}{2}\Gamma_{M} & = & A_{a},\label{eq:EqGamma}
\end{eqnarray}
where the coefficient on the right hand side is
\begin{equation}
A_{a}=\sum_{k=a+1}^{M}R_{ak}^{\textrm{pd}}+\sum_{k=1}^{M-1}R_{ka}^{\textrm{pd}}.
\end{equation}
Or equivalently, (\ref{eq:EqGamma}) can be written in a matrix form
as
\begin{equation}
B\Gamma=A
\end{equation}
with the matrix elements of $B$ given by\begin{equation} B_{jk}=\cases{ \frac{1}{2}, &\textrm{for} $k<j$,\\ \frac{1}{2}(2M-j), &\textrm{for} $k=j$,\\ 1, &\textrm{for} $j<k<M$,\\ \frac{1}{2}, &\textrm{for} otherwise.} \end{equation}Therefore,
by multiplying both sides with the inverse of $B$, we have
\begin{equation}
\Gamma=B^{-1}A.
\end{equation}

Therefore, the original CMRT master equation (cf. (\ref{eq:OriEq}))
can be written in the Lindblad form as
\begin{eqnarray}
\partial_{t}\rho & = & -\rmi[H_{e}(t),\rho]-\frac{1}{2}\sum_{k\neq k^{\prime}}R_{kk^{\prime}}^{\textrm{dis}}(t)\left[\left\{ A_{kk^{\prime}}^{\dagger}A_{kk^{\prime}},\rho\right\} -2A_{kk^{\prime}}\rho A_{kk^{\prime}}^{\dagger}\right]\nonumber \\
 &  & -\frac{1}{2}\sum_{k}\Gamma_{k}(t)\left[\left\{ A_{kk}^{\dagger}A_{kk},\rho\right\} -2A_{kk}\rho A_{kk}^{\dagger}\right],
\end{eqnarray}
as given in (\ref{eq:LCMRT}).

\section*{References}


\providecommand{\newblock}{}

\end{document}